\documentclass{article}
\usepackage[english]{babel}
\usepackage{graphicx}
\usepackage{a4wide}
\usepackage{amsmath,amssymb}
\usepackage{caption}
\usepackage{subcaption}
\usepackage{siunitx}
\usepackage{standalone}
\usepackage{pgfplots}
\usepgfplotslibrary{external} 
\usepackage{pgfplotstable}
\usepackage{tikz}
\pgfplotsset{compat=1.18}
\usepackage[acronym]{glossaries}
\usepackage{xcolor}
\usepackage{algpseudocode}
\usepackage{algorithm}
\usepackage[affil-it]{authblk}
\usepackage{multirow}

\newacronym{MILP}{MILP}{Mixed-Integer Linear Programming}
\newacronym{QAP}{QAP}{Qubit Assignment Problem}
\newacronym{QRP}{QRP}{Qubit Routing Problem}
\newacronym{QMP}{QMP}{Qubit Mapping Problem}
\newacronym{HOP}{HOP}{Heavy Output Probability}
\newacronym{KPI}{KPI}{Key Performance Indicator}

\DeclareMathOperator{\Min}{Min}
\DeclareMathOperator{\Max}{Max}
\DeclareMathOperator{\RMD}{RMD}
\newtheorem{definition}{Definition}

\title{An Exact Branch and Bound Algorithm for the generalized Qubit Mapping Problem}

\author[1]{Bjørnar Luteberget}
\author[1]{Kjell Fredrik Pettersen}
\author[1]{Giorgio Sartor}
\author[1]{Franz G. Fuchs}
\author[2]{Dominik Leib}
\author[2]{Tobias Seidel}
\author[2,3]{Raoul Heese}
\affil[1]{SINTEF Digital, Oslo, Norway}
\affil[2]{Fraunhofer ITWM, Kaiserslautern, Germany}
\affil[3]{NTT DATA, Munich, Germany}
\date{\today}

\begin{document}

\maketitle

\begin{abstract}
Quantum circuits are typically represented by a (ordered) sequence of gates over a set of virtual qubits. During compilation, the virtual qubits of the gates are assigned to the physical qubits of the underlying quantum hardware, a step often referred to as the qubit assignment problem. To ensure that the resulting circuit respects hardware connectivity constraints, additional SWAP gates are inserted as needed, which is known as  the qubit routing problem. Together, they are called the \gls{QMP}, which is known to be NP-hard. A very common way to deal with the complexity of the \gls{QMP} is to partition the sequence of gates into a sequence of gate groups (or layers). However, this imposes a couple of important restrictions: (1) SWAP gates can only be added between pairs of consecutive groups, and (2) all the gates belonging to a certain group have to be executed (in parallel) in the same time slot. The first one prevents gates to be re-arranged optimally, while the second one imposes a time discretization that practically ignores gate execution time. 
While this clearly reduces the size of the feasible space, little is still known about how much is actually lost by imposing a fixed layering when looking at the minimization of either the number of SWAPs or the makespan of the compiled circuit.
In this paper, we present a flexible branch and bound algorithm for a generalized version of the \gls{QMP} that either considers or ignores the gate layering and the gate execution time. The algorithm can find find proven optimal solutions for all variations of the \gls{QMP}, but also offers a great platform for different heuristic algorithms.
We present results on several benchmark sets of small quantum circuits, and we show how ignoring the layering can significantly improve some key performance indicators of the compiled circuit.
\end{abstract} \glsresetall

\section{Introduction}
A quantum algorithm can be viewed as applying a unitary evolution to a specified initial state and then taking measurements~\cite{NielsenChuang2010}. In the computational gate model of quantum computing, this evolution is represented by a sequence of quantum gates. Each gate implements a unitary operation on one or more qubits. Gates are usually taken from a universal set of (parameterized) gates, which enables the representation of any unitary evolution by choosing just gates from this set. To achieve universality, at least one multi-qubit gate is required because multi-qubit interactions are essential for generating entanglement.

In practice, a quantum computer supports only a fixed native gate set, which acts as the hardware-specific realization of a universal gate set. This native set usually includes several single-qubit gates and one two-qubit entangling gate. To execute an algorithm, the target unitary is compiled into a sequence of these native gates, producing a set of physically realizable instructions that can be implemented directly on the quantum device. Most hardware platforms use the controlled NOT (CX) gate as their fundamental two‐qubit operation.\footnote{Controlled‐phase gates (CZ) are also common. One has $(I \otimes \mathrm{H}) \mathrm{CZ} (I \otimes \mathrm{H}) = \mathrm{CX}$, where $\mathrm{H}$ denotes the Hadamard gate.} 
However, there is an important limitation: the application of two‐qubit gates is only possible between a fixed set of \emph{connected} pairs of physical qubits.
For example, superconducting devices often exhibit linear, star~\cite{correr2024characterizing}, or heavy‐hex connectivity~\cite{hetenyi2024creating}. Although trapped‐ion systems can support all‐to‐all connectivity~\cite{Chen2024benchmarkingtrapped}, these platforms usually remain small in qubit count before requiring the assembly of additional traps with limited connectivity.

Most of the quantum computers currently available are NISQ devices, which are inherently affected by noise and errors due to hardware imperfections. Analyzing the specific impact of this noise on the execution of quantum algorithms remains a significant challenge. However, it is clear that applying more quantum gates increases circuit depth, resulting in longer runtimes and greater exposure to noise. Therefore, a key objective is typically to either minimize the number of gates in a quantum circuit or to minimize its depth. 

Generally, the qubit connectivity of a quantum computer can be modeled as an undirected graph $\mathcal{G} = (V,E)$, where $V$ represents the set of physical qubits and $E$ is the set of edges indicating which pairs of qubits are connected. Throughout this work, we distinguish between two types of qubits:
\begin{itemize}
  \item \emph{Virtual qubits} refer to the qubits defined in the mathematical formulation of a quantum algorithm and are to be interpreted as purely abstract constructs. For this reason, they are not limited by hardware connectivity.
  \item \emph{Physical qubits} are the actual qubits a quantum computer is operating with. As such, they are subject to the limitations imposed by their specific hardware realization. One key limitation is qubit connectivity, which we model using the hardware graph $\mathcal{G}$. Another is the noise, which means that each qubit is affected by hardware-induced errors over time.
\end{itemize}

In short, virtual qubits are used in the design of quantum algorithms. In order to execute these algorithms on quantum hardware, two transformations are necessary.

\begin{enumerate}
\item \textbf{Gate Decomposition.} The quantum gates must be decomposed into the native gate set available on the quantum device. This part of the transformation is not the focus of this work. Typically, it can be realized straightforwardly, as relations between common gates are known.
A common assumption (see, for example, \cite{nannicini2022optimal}), is that such a conversion is given and the resulting set of gates consists of one-qubit gates and one two-qubit gate. In the following, we also presume (without loss of generality) that the two-qubit gate is a CX gate.
\item \textbf{Qubit Mapping.} This consists of two interconnected subproblems. The \emph{qubit assignment problem}~\cite{zhu2020exact} assigns each virtual qubit to a physical qubit\footnote{When using quantum error correction for fault-tolerant quantum computing, each virtual qubit is represented by multiple physical qubits that form a logical qubit. We do not consider this generalization here.}. The \emph{qubit routing problem}~\cite{li2019tackling} adds the necessary SWAP gates when the assignment does not satisfy the device's connectivity. In fact, when $\mathcal{G}$ is not fully connected, it becomes highly probable that any such assignment would contain at least one two-qubit gate whose virtual qubits are assigned to physical qubits that are not adjacent in $\mathcal{G}$. In this case, one or more SWAP gates need to be inserted, each of which simply swaps the virtual qubits residing on two adjecent physical qubits (at the cost of adding three CX gates). 
While the use of additional SWAP gates allows an assignment to satisfy the connectivity of the specific hardware $\mathcal{G}$, it also increases the overall error rate of quantum computations. Finding a mapping (i.e., assignment plus routing) that minimize error rate is called the \emph{\gls{QMP}}. As the error rate is difficult to compute or embed in optimization algorithms, it is typically substituted by proxy objectives such as minimizing the number of SWAP gates or minimizes the depth of the circuit. This is the main focus of this work.
\end{enumerate}




In general, \gls{QMP} is NP‐complete~\cite{siraichi2018qubit,botea2018complexity}, so most practical quantum compilers treat assignment and routing separately. Many approaches (see, for example, the well-known SABRE algorithm~\cite{li2019tackling}) start by choosing and fixing an initial assignment, and then insert SWAP gates as necessary. Note that any possible assignment admits a routing that makes the mapping feasible for the specific hardware, therefore iterative heuristic methods tend to perform well in practice (since infeasibility is not a big issue). Recent surveys such as \cite{maronese2022quantum}, \cite{ge2024quantum}, and \cite{cardama2024quantum} discuss various heuristic and exact methods for \gls{QMP}. In this work, we try to challenge some of the most common assumptions when defining the \gls{QMP} and it is important to do so via proven optimal solutions. Therefore, our focus will be on exact methods.

\begin{figure}[!ht]
    \centering
    \begin{subfigure}[c]{0.32\textwidth}
        \centering
        \includegraphics[height=0.95in]{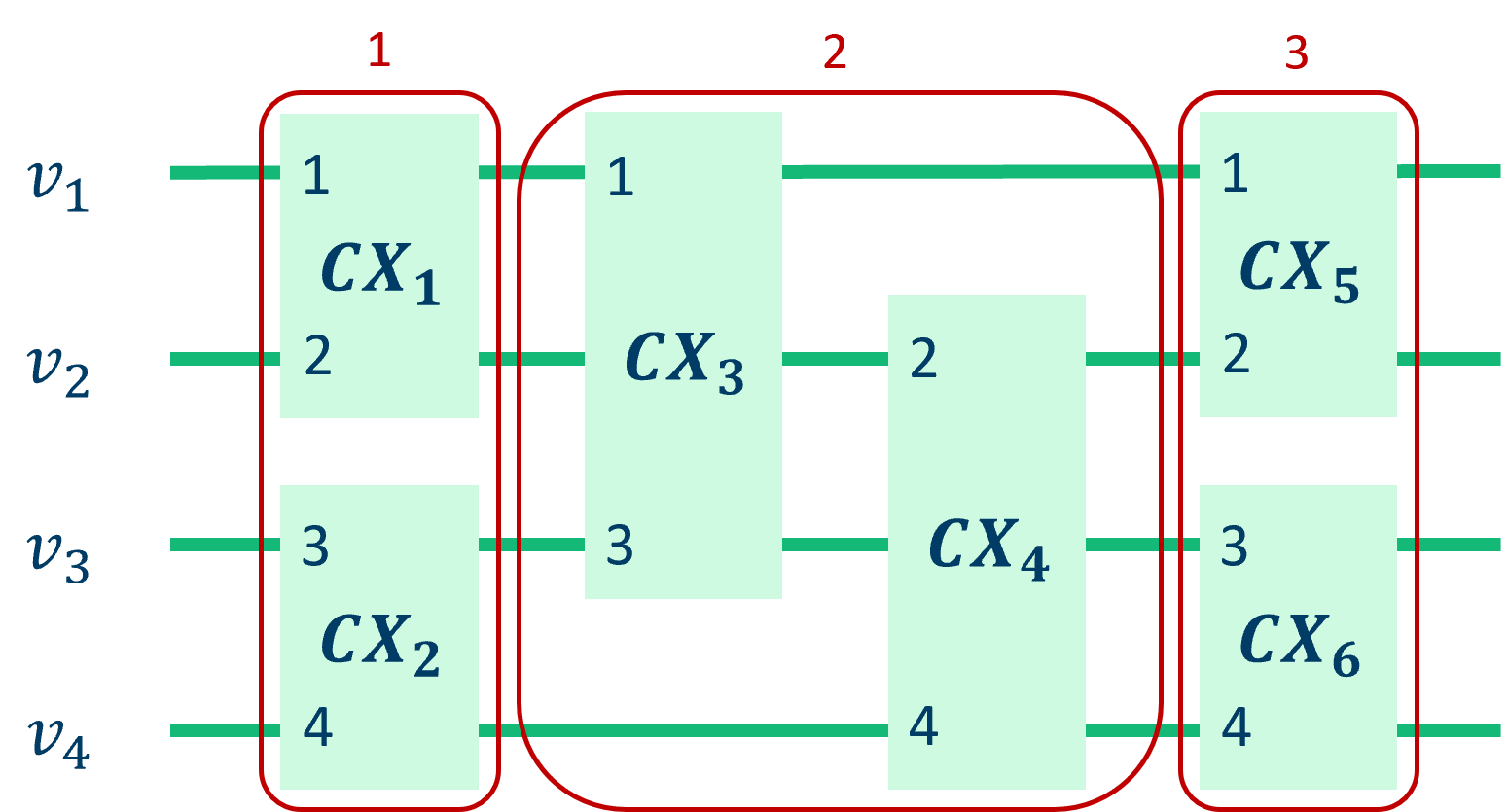}
        \caption{}
        \label{fig:virtual}
    \end{subfigure}
    \hfill
    \begin{subfigure}[c]{0.32\textwidth}
        \centering
        \includegraphics[height=0.95in]{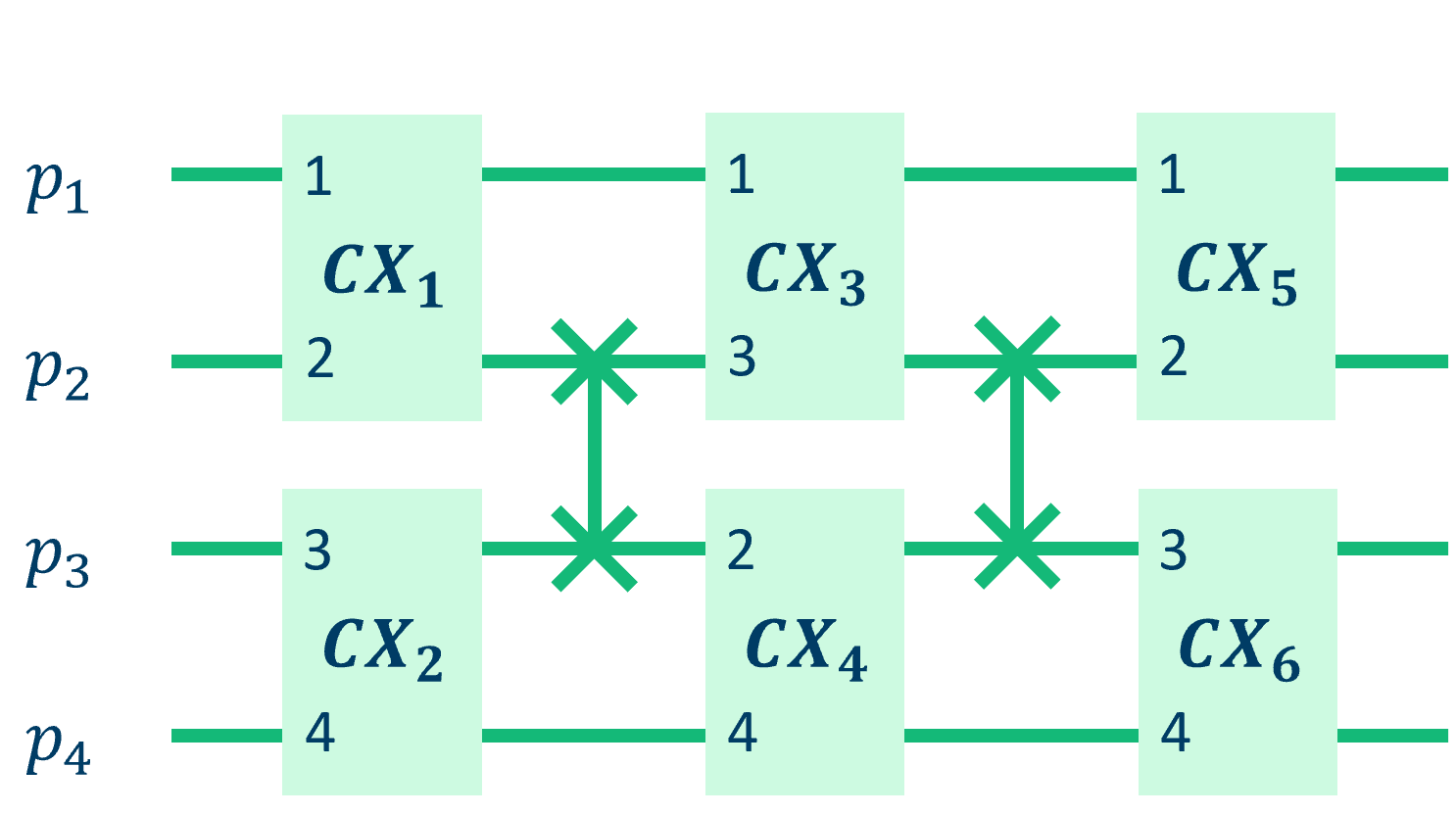}
        \caption{}
        \label{fig:physical_grouping}
    \end{subfigure}
    \hfill
    \begin{subfigure}[c]{0.32\textwidth}
        \centering
        \includegraphics[height=0.95in]{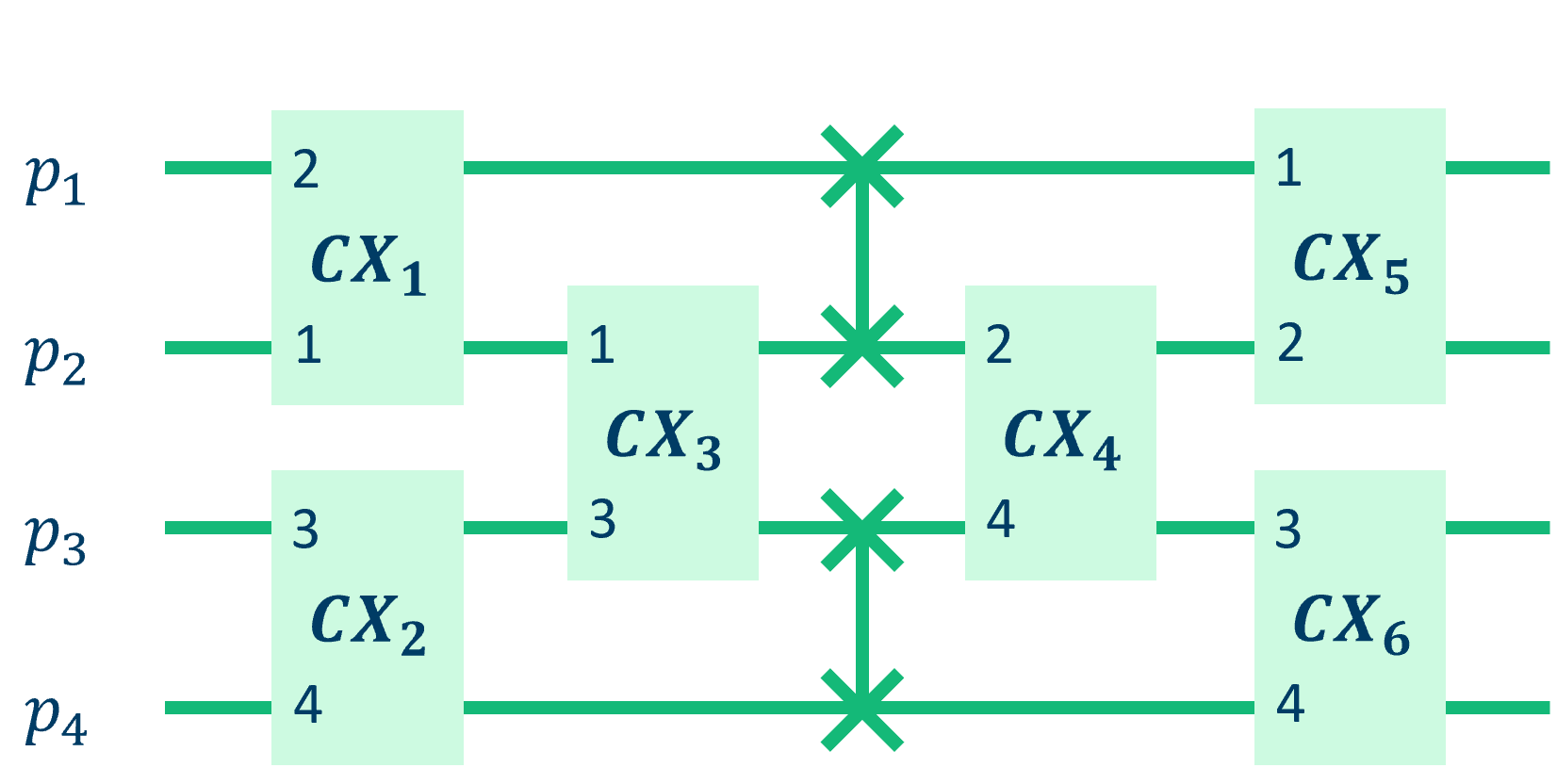}
        \caption{}
        \label{fig:physical_no_grouping}
    \end{subfigure}
    \hfill
    \caption{An example of qubit mapping with and without grouping on a linear hardware graph with four qubits $p_1 - p_2 - p_3 - p_4$, where the objective is to minimize the depth of the compiled circuit. Figure (a) shows the original circuit on virtual qubits with the 3 groups highlighted. Figure (b) shows an optimal compiled circuit with grouping, where a SWAP gate was added between each group to satisfy the hardware's connectivity. Figure (c) shows an optimal compiled circuit that ignores grouping. Since each SWAP gate requires three CX gates in series, the execution time (i.e, depth) of the circuit in (b) is equivalent to 9 CX gates, while it is only 7 for the circuit in (c).}
    \label{fig:grouping_vs_nongrouping}
\end{figure}

The literature on exact methods is rather limited and mostly involves either specific topologies (see, for example, \cite{mulderij2023polynomial}) or specific aspects of the problem (see, for example, \cite{bhattacharjee2017depth}). A notable exception is the recent work by Nannicini et al.~\cite{nannicini2022optimal}, which provides a mixed‐integer binary formulation that solves \gls{QMP} optimally for circuits up to 8 qubits. Wagner et al.~\cite{wagner2022improving} extend this by proposing a matheuristic that again treats the assignment and routing separately, achieving faster runtimes with comparable solution quality. However, both of these approaches (like many others) assume that gates are grouped into sa-called \emph{layers}, which are sets of gates that can be executed in parallel and assuming that all gates within a layer have identical duration. Between layers, one may insert SWAP gates, but those cannot occur within a layer. This assumption typically generates non-optimal mappings, as one can see in the example of Figure \ref{fig:grouping_vs_nongrouping}.
The same assumption is exploited in \cite{wille2019mapping}, where the authors choose a SAT solver to perform a mapping that minimizes the number of SWAP gates.

This assumption is instead dropped in \cite{zulehner2019compiling}, where the authors develop an $A^*$ algorithm to compile $SU(4)$ quantum circuits while again minimizing the number of SWAP gates. The layered-based approach was also dropped in \cite{zhang2021time}, in this case with a focus on time-optimal mapping, which minimizes the depth of the circuit while taking into account the execution time of each gate. Here, the authors develop a branch-and-bound algorithm that shares many ideas with ours, but employs a time-indexed approach that we think imposes an unnecessary time-discretization on the problem.

In this paper, we develop a flexible branch-and-bound algorithm that can be used to find exact solutions to a generic qubit mapping problem, that is one in which one could decide to minimize number of swaps or depth (or both at the same time with different weight) while taking into account gate execution time and deciding to impose or relax the layering constraint. This framework provides proven optimal solutions, but it can also be used as a heuristic simply by pruning more aggressively branches of the branch-and-bound tree. For example, a beam-search behavior can be achieved by limiting the number to nodes at each depth.

This flexible framework allows us to study the performance impact of several assumptions, for example by comparing solutions that satisfy the layering constraint to solutions that allow SWAP gates to be interleaved arbitrarily between two‐qubit gates.
In particular, we show that this relaxation can significantly reduce the resulting circuit depth. 


The remainder of the paper is organized as follows. In Section~\ref{sec:problem}, we formally define the \gls{QMP}. Section~\ref{sec:branchandbound} describes our branch and bound algorithm and Section~\ref{sec:experiments} presents computational results comparing our methods to existing compilers. Finally, Section~\ref{sec:conclusion} concludes and outlines future work.

\section{Problem definition}
\label{sec:problem}

\begin{figure}[!ht]
    \centering
    \begin{subfigure}[c]{0.45\textwidth}
        \centering
        \includegraphics[width=0.8\textwidth]{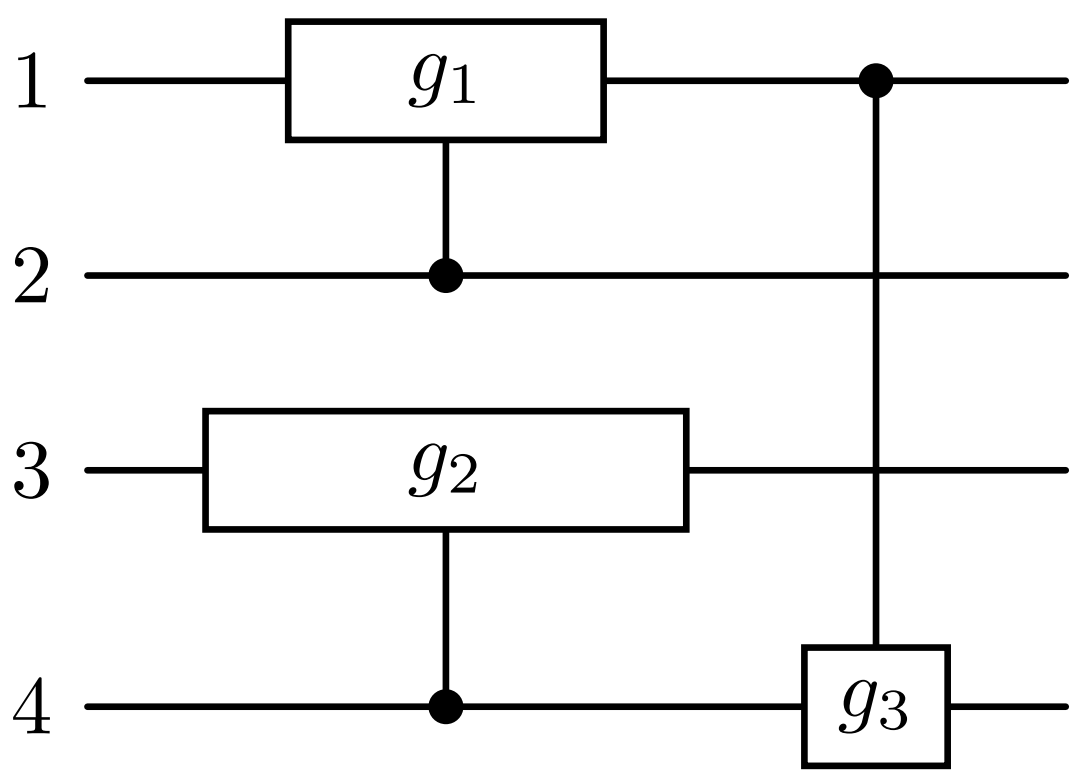}
        \caption{}
        \label{fig:simplecircuitclassic}
    \end{subfigure}
    \hfill
    \begin{subfigure}[c]{0.45\textwidth}
        \centering
        \includegraphics[width=0.8\textwidth]{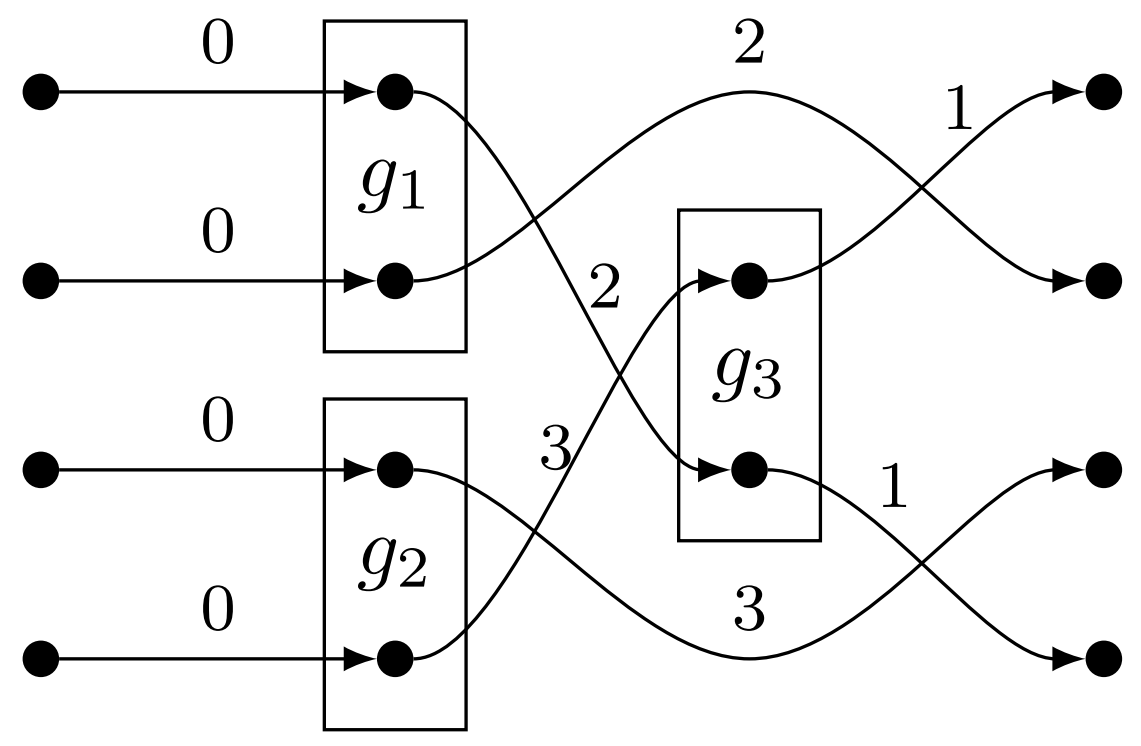}
        \caption{}
        \label{fig:simplecircuitprecedence}
    \end{subfigure}
    \hfill
    \caption{Two different ways of drawing of a simple circuit $\mathcal{C} = [\{(1,2),2\},\{(3,4),3\},\{(4,1),1\}]$, a classical circuit on virtual qubits (a) and a precedence graph (b).}
    \label{fig:simplecircuit}
\end{figure}

Let
$
    Q = \{q_1,\,q_2,\,\dots,\,q_n\}
$
be a set of virtual qubits and let
$
    G \subseteq Q \times Q
$
be the set of ordered pairs of virtual qubits on which two‐qubit gates act.
Without loss of generality, we can represent a quantum circuit as a sequence
\begin{equation}
    \label{eq:circuit}
    \mathcal{C} = \left[\{g_1,d_1\},\{g_2,d_2\},\dots,\{g_N,d_N\}\right],\quad
    g_i = (p_i,q_i) \in G,\ p_i,q_i \in Q,\ p_i \neq q_i,\ d_i \geq 0,
\end{equation}
where $g_i$ denotes a two‐qubit gate acting on the (ordered) pair of virtual qubits $(p_i, q_i)$ (although the order of $(p_i, q_i)$ can often be ignored, the ordering simplifies our notation), and $d_i$ is the duration (i.e., execution time) of gate $g_i$.
As noted in \cite{zulehner2019compiling}, a circuit implicitly induces a partial ordering of gates. For example, consider the circuit $\mathcal{C} = [\{(1,2),2\},\{(3,4),3\},\{(4,1),1\}]$. This can be drawn simply as a classical circuit on virtual qubits (see Figure \ref{fig:simplecircuitclassic}) or as a directed graph of precedence relationships (see Figure \ref{fig:simplecircuitprecedence}). In particular, Figure \ref{fig:simplecircuitprecedence} displays the flow of information through the quantum circuit with a so-called precedence graph (which is commonly used in scheduling applications) and highlights the precedence relationships between pairs of gates. In this graph, an arc of length $l$ between two nodes indicates that the event associated to the target node can only start $l$ time units after the start of the event associated with the source node. In our case, the length of the arcs represent gate execution time. These times can then be used to track the total execution time of the compiled circuit and to determine the minimum time after which a gate can be executed. For example, if $g_1$ and $g_2$ are executed at the same time, then $g_3$ can be executed only after $\max(2,3) = 3$ time units. Tracking execution time precisely can be important, as the execution time of a $CX$ gate is typically 10 times the one of a single-qubit gate, and a SWAP gate is 3 times longer (as it combines 3 $CX$ gates in series). Although this graph will not be mentioned explicitly, these precedence relationships will be exploited by the algorithm described in the next section to drastically reduce the search space and guarantee feasibility.
Given these considerations, one could in fact see the qubit mapping problem as a scheduling problem, one in which the gates need to be \emph{scheduled} in a physical quantum hardware while satisfying the connectivity constraints and minimizing the total execution time. We can then introduce a formal definition for our interpretation of the qubit mapping problem. 
\begin{definition}
\label{def:qsp}
Given a set of virtual qubits $Q$, a circuit $\mathcal{C}=[\{g_1,d_1\}, \dots, \{g_N,d_N\}]$ as described in \eqref{eq:circuit}, and a hardware graph $\mathcal{G}=(V,E)$, the \Acrfull{QMP} consists of finding a schedule $\mathcal{S}=[\{a_1,t_1\}, \dots, \{a_M,t_M\}]$ such that each $a_k$, $k\in\{1,\dots,M\},\ M \geq N$, is an ordered assignment $(i,j)$, $i,j\in V$ of physical qubits, and $t_k$ is its scheduled time of execution $t_k\geq0$. In other words, the schedule determines over which physical qubits a gate should be executed and at which time. Some $a_k$ are associated to the original gates $g_1,\dots,g_N$ while others will be associated to the necessary SWAP gates. Such a schedule $\mathcal{S}$ is considered feasible if:
\begin{itemize}
    \item (Assignment constraints) each assignment $(i,j)$ belongs to the set of edges $E$;
    \item (Routing constraints) the ``flow'' of virtual qubits through the hardware must satisfy all precedence relationships defined by the circuit $\mathcal{C}$, and whenever a gate $(p,q)$, $p,q\in Q$ is assigned to the physical qubits $(i,j)$, $i,j\in V$, then $i,j$ must ``contain'' the virtual qubits $p,q$ at the time of execution;
\end{itemize} A feasible schedule is optimal when it minimizes the total execution time of the circuit.
\end{definition}

Note that a schedule $\mathcal{S}$ implicitly determines the initial assignment of virtual qubits to physical qubits, simply by looking the at the virtual qubits of the first gate that is ever applied to those physical qubits. For example, if the first gate executed on the physical qubits $i=1,j=2$ of a certain hardware graph was operating on the virtual qubits $p=3,q=4$, then we determine the initial assignment to be $i\gets p$ and $j\gets q$.

\subsection{An example of a feasible schedule}
\label{sec:scheduleexample}
To highlight the role of the assignment and routing constraints, consider the circuit of Figure \ref{fig:simplecircuit} and the linear hardware graph $1 - 2 - 3 - 4$.
(note that if the hardware graph was a complete graph with at least 4 nodes, then the problem would become trivial). Then, consider the na\"{i}ve schedule $\mathcal{S}_1 = [\{(1,2),0\}_{g_1},\{(3,4),0\}_{g_2},\{(4,1),3\}_{g_3}]$ (here the subscript is simply used to indicate which gate a certain qubit assignment is associated with). This schedule clearly does not satisfy the hardware graph constraints, since the edge $(4,1)$ (or $(1,4)$) does not exist in the hardware graph. The schedule $\mathcal{S}_2 = [\{(1,2),0\}_{g_1},\{(3,4),0\}_{g_2},\{(3,4),3\}_{g_3}]$ satisfies all hardware graph constraints, but does not satisfy the routing constraints as the physical qubits 3 and 4 do not contain the correct virtual qubits 1 and 4. Then, we can add a pair of swap gates $s_1=(1,2)$ and $s_2=(1,3)$ to first swap virtual qubit 1 with 2 and then again 1 and 3, so that virtual 1 resides on physical qubit 3 before the execution of gate $g_3$. Assuming each swap gate takes 6 time units, a feasible schedule would be $\mathcal{S}_3 = [\{(1,2),0\}_{g_1},\{(3,4),0\}_{g_2},\{(1,2),2\}_{s_1},\{(2,3),8\}_{s_2},\{(3,4),14\}_{g_3}]$.

Note that this is far from being optimal. In fact, there exists a schedule that does not even require additional swap gates, that is $\mathcal{S}_4 = [\{(2,1),0\}_{g_1},\{(4,1),0\}_{g_2},\{(2,3),3\}_{g_3}]$.


\section{A branch and bound algorithm}
\label{sec:branchandbound}
To solve the \gls{QMP} we consider an exact branch and bound algorithm as it offers great flexibility both in the modeling and in the  search strategies. In fact, with this framework it is easy to switch between different objective functions or to heuristically limit the search space if needed. A typical branch and bound algorithm is a tree search algorithm that recursively breaks the original minimization problem in smaller subproblems until it can prove that either a certain subproblem is optimal or the original problem is infeasible. An iteration of the algorithm consists of the following steps:
\begin{enumerate}
    \item \textit{Selection:} Choose a subproblem from the list of open subproblems and remove it from the list. If the list is empty, terminate.
    \item \textit{Bounding/Fathoming:} Compute a lower bound for the selected subproblem. If the subproblem is infeasible or the lower bound is greater than the best known feasible solution, go to 1.
    \item \textit{Branching/Expansion:} If the solution of the subproblem is feasible for the original problem, update the best known feasible solution. Otherwise, split the subproblem in two or more (disjoint) subproblems and add them to the queue of open subproblems.
\end{enumerate}

In the rest of the section, we describe in details how we perform all this steps. The algorithm explores state nodes, starting with a root node $n_0$ with no scheduled gates, and expanding leaf nodes by adding unscheduled circuit gates or SWAP gates. To avoid exploring redundant and sub-optimal branches in the search tree, we will only expand nodes that belong to a Pareto front of all nodes added so far. The relevant part of the search tree in the algorithm can be presented as
\begin{equation}
    \label{eq:searchtree}
    T = \{T_P,T_U\}
\end{equation}
where $T_P$ is the set of all Pareto front nodes in the search tree, while $T_U\subseteq T_P$ are the currently unexpanded nodes. The overall tree search algorithm is described in Algorithm~\ref{alg:treesearch}.

\begin{algorithm}
    \caption{TreeSearch}
    \begin{algorithmic}
    \Require Virtual qubits $Q$, circuit $\mathcal{C}$ as described in \eqref{eq:circuit} and hardware graph $\mathcal{G} = (V,E)$
    \Ensure $\mathcal{S}=[\{a_1,t_1\}, \dots, \{a_N,t_N\}]$ solution to the qubit mapping problem for $Q$, $\mathcal{C}$ and $\mathcal{G}$.
    \State $T_P,T_U \gets \{n_0\}$\Comment{Initialize the search tree}
    \While{$\# T_U > 0$}
        \State $n \gets n\in T_U\text{ with minimal }h(n)$\Comment{Node to expand}
        \State Remove $n$ from $T_U$
        \If{$n$ schedules all gates in $\mathcal{C}$}
            \State \Return{Solution by parent backtracking from $n$}
        \Else
            \State Expand($n$)
        \EndIf
    \EndWhile
    \State No solution found
    \end{algorithmic}
    \label{alg:treesearch}
\end{algorithm}

A node $n\in T_N$ in the search tree different from the root node $n_0$ can be represented as
\begin{equation}
    \label{eq:treenode}
    n = (f_n,i_n,e_n)
\end{equation}
where $f_n$ is the parent node of $n$, $i_n\in\{0,\dots,N\}$ is the index of the gate scheduled from $f_n$ to $n$, either a SWAP gate if $i_n=0$ or a circuit gate $\{g_{i_n},d_{i_n}\}\in\mathcal{C}$ if $i_n>0$, and $e_n=(v_n,w_n)\in E$ is the hardware position where the gate was applied. By parent node backtracking, the node $n$ implicitly defines the set $\mathcal{C}_n\subseteq \mathcal{C}$ of the problem circuit gates that have not been scheduled yet and the set $Q_n\subseteq Q$ of all virtual qubits with at least one scheduled gate in the parent chain up to $n$. The set $Q_n$ contains all qubits with at least one gate in $\mathcal{C}\setminus\mathcal{C}_n$, but may also contain additional qubits from scheduled SWAP gates.

Each node defines the assignment map $A_n:Q_n\rightarrow V$ mapping virtual qubits to their current hardware positions. The node expansions must ensure consistency and injectivity of the assignment map. When new qubits are added to $Q_n$, they must first be assigned to an unassigned hardware position in the parent node, while the rest of the assignment map must be the same as for the parent. The positions $v_n$ and $w_n$ must be the assigned positions of the virtual qubits the gate is applied for, and in the case of a SWAP gate the assignment map must swap the values for the two gate qubits.

The depth $D_n(v)$ is the earliest time after the execution of all gates on a hardware qubit $v\in V$. It will be the scheduled time of $g$ plus the duration of $g$ where $g$ is the last gate scheduled to be executed on an edge of $v$, or 0 if no such gate has been scheduled. Gates are scheduled as early as possible after the completion of the previous gates on any of the same hardware qubits. The map $D_n$ defines a depth map $D^Q_n$ on $Q$ where $D^Q_n(q)=D_n(A_n(q))$ for assigned qubits $q\in Q_n$, and $D^Q_n(q)=0$ for unassigned qubits.

\subsection{Node expansion}
When a node $n$ is expanded, the algorithm will try to add child nodes to $n$ by adding new gates to the schedule without getting in conflict with the existing assignment. Both problem circuit gates and SWAP gates may be added. A circuit gate $g_i\in\mathcal{C}_n$ can only be added if it is minimal, i.e. the first unscheduled gate for both of its qubits. The gate must be added to a hardware edge and the two virtual qubits must be assigned to the two nodes of the hardware edge so that a qubit $q\in Q_n$ is assigned to $A_n(q)$ and a qubit $q\in Q\setminus Q_n$ is assigned to a hardware node $v\notin A_n(Q_n)$. A SWAP gate can be added to any hardware edge and will swap the assignment values in $A_n$ for the gates already assigned to any of the edge positions, though our implementation does not add SWAPs to edges where neither position has a virtual qubit assigned to it. The expansion algorithm is described in Algorithm~\ref{alg:expansion}.

The algorithm uses a lazy assignment; a virtual qubit node is not assigned to a hardware position before it is first assigned by a gate. In this way, the algorithm solves both the routing and the initial qubit assignment, which is built through parent backtracking in the solution.

\begin{algorithm}
    \caption{Expand}
    \begin{algorithmic}
    \Require Search tree $T$ as described in \eqref{eq:searchtree} and an unexpanded node $n$ as described in \eqref{eq:treenode}
    \Ensure Children nodes of $n$ that may lead to the final solution are added to the search tree.
    \ForAll{$(g_i=(p_i,q_i),d_i)$ minimal element in $\mathcal{C}_n$}\Comment{Try schedule circuit gates}
        \If{$p_i,q_i\in Q_n$}
            \If{$(A_n(p_i),A_n(q_i))\in E$}
                \State $n'\gets (n,i,(A_n(p_i),A_n(q_i)))$
                \State TryInsert$(T,n')$
            \EndIf
        \ElsIf{$p_i\in Q_n$ and $q_i\notin Q_n$}
            \ForAll{$w\in V\setminus A_n(Q_n)$ such that $(A_n(p_i),w)\in E$}
                \State $n'\gets (n,i,(A_n(p_i),w))$
                \State TryInsert$(T,n')$
            \EndFor
        \ElsIf{$p_i\notin Q_n$ and $q_i\in Q_n$}
            \State{Similar to $p_i\in Q_n$ and $q_i\notin Q_n$}
        \Else\Comment{$p_i,q_i\notin Q_n$}
            \ForAll{$(v,w)\in E$ such that $v,w\notin A_n(Q_n)$}
                \State $n'\gets (n,i,(v,w))$
                \State TryInsert$(T,n')$
            \EndFor
        \EndIf
    \EndFor
    \ForAll{$(v,w)\in E$ such that $v\in A_n(Q_n)$ or $w\in A_n(Q_n)$}\Comment{Try schedule SWAP gates}
        \State $n'\gets (n,0,(v,w))$
        \State TryInsert$(T,n')$
    \EndFor
    \end{algorithmic}
    \label{alg:expansion}
\end{algorithm}

\subsection{The Pareto front}
The tree search nodes can be grouped into equivalence classes according to their state. Two nodes $n_1$ and $n_2$ are equivalent if they have the same assignment of virtual qubits to the hardware positions, and the same set of unscheduled gates, i.e. $Q_{n_1}=Q_{n_2}$, $A_{n_1}=A_{n_2}$, and $\mathcal{C}_{n_1}=\mathcal{C}_{n_2}$. The Pareto front nodes $T_P$ are the nodes that are part of the Pareto front of minimal depth values for their states. So when we create a new node $n$ in the algorithm, it can be pruned if there is already a node $n'\in T_P$ with the same state as $n$, and where $D_{n'}$ is equal to or dominated by $D_n$. Otherwise $n$ will be added to $T_P$ and $T_U$, while other nodes that are no longer in the Pareto front will be removed from $T_P$ and $T_U$, as described in Algorithm~\ref{alg:tryinsert}.

\begin{algorithm}
    \caption{TryInsert}
    \begin{algorithmic}
    \Require Search tree $T$ as described in \eqref{eq:searchtree} and node $n$ as described in \eqref{eq:treenode}
    \Ensure $n$ is added to the search tree if it represents a new Pareto front element.
    \If{there is no $n'\in T_P$ with same state as $n$ such that $D_{n'}\ll D_n$}
        \ForAll{$n'\in T_P$ with same state as $n$}
            \If{$D_{n}\ll D_n'$}
                \State Remove $n'$ from $T_P$
                \If{$n'\in T_U$}
                    \State Remove $n'$ from $T_U$
                \EndIf
            \EndIf
        \EndFor
        \State Add $n$ to $T_P$ and $T_U$
    \EndIf
    \end{algorithmic}
    \label{alg:tryinsert}
\end{algorithm}

\subsection{Lower bound function}
To help the algorithm search for an optimal solution to the qubit mapping problem, we use a measure $h(n)$ on the total execution time of the circuit for the best solution from $n$. The tree search algorithm uses this measure by always expanding a node from $T_U$ that minimizes $h(n)$. If $h(n)$ is always a lower bound, the search algorithm will be admissible and always return an optimal solution if solutions exist.

To help us define $h(n)$ as a lower bound, we need some utility measures on the circuit gates.

\begin{definition}
Given a set of virtual qubits $Q$ and a circuit $\mathcal{C}=[\{g_1,d_1\}, \dots, \{g_N,d_N\}]$ as described in \eqref{eq:circuit}.

The set $\mathcal{C}^q$ is the set of all gates that apply to the qubit $q$.
\begin{equation}
    \mathcal{C}^q=[\{g_i,d_i\}: g_i=(p_i,q_i),\ q\in\{p_i,q_i\}],\quad q\in Q
\end{equation}

The set $\mathcal{C}^{q,i}$ is the set of all gates that apply to the qubit $q$ from $g_i$ or later.
\begin{equation}
    \mathcal{C}^{q,i}=[\{g_j,d_j\}: \{g_j,d_j\}\in\mathcal{C}^q,\ j\geq i\}],\quad q\in Q,\ i\in\{1,\dots,N\}
\end{equation}

The minimal time $\delta(i)$ from the scheduled time of $g_i$ to the scheduled completion of the entire circuit is recursively given by
\begin{equation}
    \delta(i) = d_i+\Max(0,\{\delta(j)\ \forall j:\{g_i,d_i\}<\{g_j,d_j\}),\quad i\in\{1,\dots, N\}
\end{equation}
where the relation $\{g_i,d_i\}<\{g_j,d_j\}$ expresses the partial ordering of the gates in the circuit.

The remaining execution time $\lambda(q,i)$ of gates that apply to the qubit $q$ from $g_i$ or later is given by
\begin{equation}
    \lambda(q,i) = \sum_{\{g_j,d_j\}\in\mathcal{C}^{q,i}}d_j,\quad q\in Q,\ \{g_i,d_i\}\in\mathcal{C}^q
\end{equation}
\end{definition}

From a node in the search tree, the depth of a qubit in a final solution cannot be smaller than its current depth plus the minimal time from the scheduled time of its first unplanned gate (if any) to the scheduled time of the entire circuit. This gives a first simple lower bound on the optimal solution depth from a node, which is the actual depth if the remaining gates were to be planned on a complete hardware graph.

\begin{definition}
Given a set of virtual qubits $Q$, a circuit $\mathcal{C}=[\{g_1,d_1\}, \dots, \{g_N,d_N\}]$ as described in \eqref{eq:circuit} and a node $n$ in the search tree.

The set $Q^U_n$ is the set of virtual cubits that still have unscheduled gates in $\mathcal{C}_n$
\begin{equation}
    Q^U_n = \{q\in Q: \mathcal{C}^q\cap \mathcal{C}_n\neq\emptyset\}
\end{equation}

For a virtual qubit $q$ with unscheduled gates, we define $m^q_n$ to be the index of the first such gate.
\begin{equation}
    \{g_{m^q_n},d_{m^q_n}\} = \Min(\mathcal{C}^q\cap \mathcal{C}_n),\quad q\in Q^U_n
\end{equation}

The qubit based depth heuristic $h^Q$ on the node $n$ is defined by
\begin{equation}
      h^Q(n) = \Max \left(\left\{ D^Q_n(q) + \delta(m^q_n) \right\}_{q\in Q^U_n} \bigcup \left\{ D^Q_n(q) \right\}_{q\in Q\setminus Q^U_n} \right)
\end{equation}
\end{definition}

To improve this heuristic, we include an estimate on the possible delays from repositioning qubits of unscheduled gates. For two hardware qubits $v$ and $w$ we let $\Pi(v,w)$ be all minimal hardware paths $\pi=(v^1_{\pi},\dots,v^{N_{\pi}}_{\pi})$ where $v^1_{\pi}=v$, $v^{N_{\pi}}_{\pi}=w$ and $(v^i_{\pi},v^{i+1}_{\pi})\in E$ for all $i$. Minimality means no subset of hardware qubits can be removed from $\pi$ and still give a valid hardware path from $v$ to $w$.

If we try by SWAP operations to move the virtual qubit $p$ from position 1 to position $j$ on the hardware path $\pi$ before we execute the gate $g_i$ in $\mathcal{C}^p$, there are some lower bounds on how early the gate can be scheduled. One is based on the duration of the gates that must be applied on $p$ before $g_i$, this is $\lambda(p,m^p_n)-\lambda(p,i)$ for the circuit gates plus $(j-1)d^S$ for the SWAP gates. The other is based on the depths $D_n(v^k_{\pi})$ of the hardware qubits that swap with $p$ for $k\in\{2,\dots,j\}$. A similar set of bounds exists for the virtual qubit $q$ to be moved from position $N_{\pi}$ to position $j+1$ on $\pi$. This gives the following scheduling bound for an unscheduled gate by moving its qubits to a specific path edge.

\begin{definition}
Given a set of virtual qubits $Q$, a circuit $\mathcal{C}$ as described in \eqref{eq:circuit}, a node $n$ as described in \eqref{eq:treenode}, an unscheduled gate $g_i=(p_i,q_i)$ such that $\{g_i,d_i\}\in \mathcal{C}_n$, a path $\pi\in \Pi(A_n(p_i),A_n(q_i))$ and a position $j\in\{1,\dots,N_{\pi}-1\}$. Then we define
\begin{equation} 
\begin{split}
    H_n(i,\pi,j)=\Max \Big( \big\{ & D^Q_n(p_i)+\lambda(p_i,m^{p_i}_n)-\lambda(p_i,i)+(j-1)d^S, \\
        & D^Q_n(q_i)+\lambda(q_i,m^{q_i}_n)-\lambda(q_i,i)+(N_{\pi}-j-1)d^S\big\} \\
    \bigcup \big\{ & D_n(v^k_{\pi})+(j+1-k)d^S\big\}_{k\in\{2,\dots,j\}} \\
    \bigcup \big\{ & D_n(v^k_{\pi})+(k-j)d^S\big\}_{k\in\{j+1,\dots,N_{\pi}-1\}}\Big)
\end{split}
\end{equation}
which is a lower bound on how early $g_i$ can be scheduled by moving $p_i$ and $q_i$ along $\pi$ to the edge $(v^j_{\pi},v^{j+1}_{\pi})$.
\end{definition}

By taking the minimum of $H_n(i,\pi,j)$ over all minimal paths $\pi$ between the hardware positions of $p_i$ to $q_i$, and all edge positions $j$ in the hardware graph, we get a lower bound on the scheduled start time of the gate $g_i$. By adding $\delta(i)$ we get the lower bound contribution from $g_i$ to the heuristic $h^G(n)$, which together with $h^Q(n)$ defines the heuristic $h(n)$ used in the tree search.

\begin{definition}
Given a set of virtual qubits $Q$, a circuit $\mathcal{C}=[\{g_1,d_1\}, \dots, \{g_N,d_N\}]$ as described in \eqref{eq:circuit} and a node $n$ as described in \eqref{eq:treenode}. The gate based depth heuristic $h^G$ on the node $n$ is defined by
\begin{equation}
      h^G(n) = \Max \{\delta(i) + \Min\{ H_n(i,\pi,j):\ \pi\in\Pi(A_n(p_i),A_n(q_i)),\ j\in\{1,\dots,N_{\pi}-1\} \}
\end{equation}
taken over all gates $g_i=(p_i,q_i)$ in $\mathcal{C}_n$ such that $p_i,q_i\in Q_n$.

The function $h_D$ giving a lower bound for the depth of optimal solution from $n$ is defined by
\begin{equation}
    \label{eq:lowerBoundDepth}
      h_D(n) = \Max \{h^Q(n), h^G(n)\}
\end{equation}
\end{definition}
which is also our lower bound function $h$ used by the algorithm to select the next node to expand.
Notice that to get $h^G$, it is sufficient to look at the gates that are the first unscheduled common gate of the two qubits they apply to, i.e. all $g_i=(p_i,q_i)$ where $\{g_i,d_i\}=\Min(\mathcal{C}^{p_i}\cap\mathcal{C}^{q_i}\cap\mathcal{C}_n)$.

\subsection{Consider SWAP count in the objective}
Instead of searching for a solution that minimizes the qubit depths, we can let the objective be to minimize the number of SWAP gates added to the solution, or a weighted combination of depth and SWAP gate count. The Pareto front value domain in the algorithm will must then be extended with an extra dimension for $s_n$, the number of scheduled SWAP gates.

As a lower bound $h_S$ for the total number of SWAPs in the final solution, we look for the unscheduled gate with longest distance between the current position of its qubits, and add the number of SWAPs needed to position the qubits next to each other. This gives
\begin{equation}
      \label{eq:lowerBoundSwap}
      h_S(n) = s_n + \Max \{ d_{\mathcal{G}}(A_n(p_i),A_n(q_i))-1 \}
\end{equation}
taken over all gates $g_i=(p_i,q_i)$ in $\mathcal{C}_n$ such that $p_i,q_i\in Q_n$, where $d_{\mathcal{G}}$ is the distance function on the hardware graph.

The combined lower bound function $h$ is updated to combine measures on depth and SWAPs by
\begin{equation}
      \label{eq:lowerBoundDepthAndSwap}
      h(n) = w_D h_D(n) + w_S h_S(n)
\end{equation}
where $w_D$ and $w_S$ are the weights of the depth and number of SWAPs in the objective function.



\subsection{Layering of gates}
The algorithm also allows us to impose the layering constraints. In fact, with the following modification, the algorithm basically follows the same constraints described in \cite{nannicini2022optimal}.

Each gate $\{g_i,d_i\}$ can be given a layer index $L(i)$, recursively defined by 
\begin{equation}
    L(i) = \Max(0,\{1+L(j)\ \forall j:\{g_j,d_j\}<\{g_i,d_i\}),\quad i\in\{1,\dots, N\}.
\end{equation}
The gates with layer $0$ are all gates that can be planned before any other gate, the gates with layer $1$ are the remaining gates that can
be planned if all gates of layer $0$ are planned, etc.

It is possible to run the algorithm in layering mode. This would imply that the node expansion in Algorithm~\ref{alg:expansion} will not add any node that schedules a gate with layer $l$ before all gates with layer $l-1$ have been scheduled. If we run the search algorithm in layering mode, we may improve the execution time since the search tree is narrowed, but there is a risk that we can miss an optimal solution where a gate must be planned before one with a lower layer.




\section{Computational results}
\label{sec:experiments}

We compare the layered to the non-layered approach by computing solutions for both methods with Algorithm \ref{alg:treesearch} and comparing the results with metrics introduced in the following section. We use different hardware graphs to highlight the effect on the metrics used when there is a large difference in connectivity and centrality of the graph. Finally, we add noise to the results and compute the heavy output probability, a metric widely used in analysis of gate-based scheduling operations (see for example \cite{PhysRevA.100.032328} and \cite{10.5555/3135595.3135617}). All implementations have been done in Python.

\subsection{Instance generation and metrics}

For each instance we create a randomly generated circuit, using the top five gates displayed in Table \ref{tab:gates}, which are derived from the IBM \texttt{ibm\_kyoto} backend. To determine the timings, the gates were transpiled as single-gate circuits by Qiskit and their duration was measured. The gate durations of Table \ref{tab:gates} are those measured timings, divided by $16$. Additionally we measured the time for a SWAP operation, which is needed for transpilation but not explicitly used for initial circuit generation and also divide it by $16$. Single-qubit gates are not scheduled explicitly by Algorithm \ref{alg:treesearch}, but contribute to the time of the two-qubit gate acting on the same qubit. To do this, we count each single-qubit gate up to the next two-qubit gate in the circuit and add their durations to the scheduling time of the latter. For simplicity, the last single-qubit gates in the circuit, if existing, are omitted when computing the circuit duration.

\begin{table}[ht]
    \centering
    \begin{tabular}{c|c|c}
      Abbrevation & Gate & Gate Duration \\ \hline
      ECR   &  Echoed Cross Resonance & 4\\
      ID   &  Identity  & 1\\
      RZ & $z$-Rotation & 1\\
      SX & Square Root & 1\\
      X & $x$-Rotation & 1\\
      \hline 
      SWAP & Swap Qubits  & 15
    \end{tabular}
    \caption{Gates used for circuit generation.}
    \label{tab:gates}
\end{table}

\begin{figure}[!ht]
    \centering
    \begin{subfigure}[c]{0.32\textwidth}
        \centering
        \begin{tikzpicture}[scale=0.5]
    \filldraw[fill=black] (0, 0) circle (4pt);
    \filldraw[fill=black] (1.5, 0) circle (4pt);
    \filldraw[fill=black] (3, 0) circle (4pt);
    \filldraw[fill=black] (4.5, 0) circle (4pt);
    \filldraw[fill=black] (6, 0) circle (4pt);
    \filldraw[fill=black] (7.5, 0) circle (4pt);
    
    \draw (0, 0) -- (1.5, 0);
    \draw (1.5, 0) -- (3, 0);
    \draw (3, 0) -- (4.5, 0);
    \draw (4.5, 0) -- (6, 0);
    \draw (6, 0) -- (7.5, 0);
\end{tikzpicture}
        \caption{Linear graph.}
        \label{fig:lineargraph}
    \end{subfigure}
    \hfill
    \begin{subfigure}[c]{0.32\textwidth}
        \centering
        \begin{tikzpicture}[scale=0.5]
    \filldraw[fill=black] (0, 0) circle (4pt) {};
    \filldraw[fill=black] (1.5, 0) circle (4pt) {};
    \filldraw[fill=black] (3, 0) circle (4pt)  {};
    
    \filldraw[fill=black] (0, 1.5) circle (4pt) {};
    \filldraw[fill=black] (1.5, 1.5) circle (4pt) {};
    \filldraw[fill=black] (3, 1.5) circle (4pt) {};
    
    \draw (0, 0) -- (1.5, 0);
    \draw (1.5, 0) -- (3, 0);
    \draw (0, 1.5) -- (1.5, 1.5);
    \draw (1.5, 1.5) -- (3, 1.5);
    \draw (0, 0) -- (0, 1.5);
    \draw (1.5, 0) -- (1.5, 1.5);
    \draw (3, 0) -- (3, 1.5);
\end{tikzpicture}
        \caption{Grid graph.}
        \label{fig:gridgraph}
    \end{subfigure}
    \hfill
        \begin{subfigure}[c]{0.32\textwidth}
        \centering
        \begin{tikzpicture}[scale=0.5]
    \filldraw[fill=black] (0, 0) circle (4pt); 
    \filldraw[fill=black] (-1.5, 2) circle (4pt); 
    \filldraw[fill=black] (-0.75, 1) circle (4pt); 
    \filldraw[fill=black] (1.5, 2) circle (4pt); 
    \filldraw[fill=black] (0.75, 1) circle (4pt); 
    \filldraw[fill=black] (0, -1.1) circle (4pt); 

    \draw (0, 0) -- (-1.5, 2);
    \draw (0, 0) -- (1.5, 2);
    \draw (0, 0) -- (0, -1.1);
\end{tikzpicture}
        \caption{Y graph.}
        \label{fig:ygraph}
    \end{subfigure}
    \hfill
    \caption{Exemplary representations of the three hardware graphs used, with six physical qubits each.}
    \label{fig:hardwaregraphs}
\end{figure}
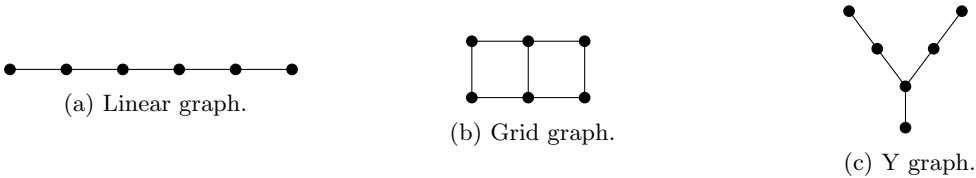

We compare the solutions of Algorithm \ref{alg:treesearch} with layering on and off on three hardware graphs with a varying number of qubits, a Linear graph, a Grid graph and a Y graph, see Figure \ref{fig:hardwaregraphs}. Each hardware graph used has unique specifications regarding connectivity and centrality which will reflect in the solution quality.

We generate $100$ instances for each combination of number of qubits in the hardware graph and a \emph{depth} parameter, which is used by Qiskit for random circuit generation. For generation we make use of the \texttt{random\_circuit()} function provided by the \texttt{qiskit.circuit.random} module. The gates scheduled by Qiskit for the circuit generation are all available gates; we transpile the output circuit again on a complete hardware-graph with the backend used to achieve a circuit that only consists of the top five gates from Table \ref{tab:gates}. Through the equivalent conversion of gates, the depth of the circuit changes.

\begin{table}[!ht]
    \centering
    \begin{tabular}{c|c|c|c|c}
      Hardware graph & Num qubits & Depth parameter & Num instances total\\ \hline
      Linear   &  4, 5, 6 & 10, 15, 20 & 900\\
       Grid & 4, 6 & 10, 15, 20 & 600\\
       Y & 4, 5, 6 & 10, 15, 20 & 900 \\
    \end{tabular}
    \caption{Instances}
    \label{tab:instances}
\end{table}

The instances are solved using a Python implementation of Algorithm \ref{alg:treesearch}, which is based on two objective functions: the number of SWAPs and the circuit depth, with the latter weighted by the durations listed in Table \ref{tab:gates}. For each objective we run Algorithm \ref{alg:treesearch} two times, one time with layering being active and one time being inactive (non-layered). The solutions are then compared with three metrics; the depth, the number of SWAPs and additionally the (unweighted) depth, which is the makespan of the circuit w.r.t. the number of gates.  Additionally we compare for both objectives the \emph{heavy output probability} of the distributions of the solutions, which is the probability that a noisy repetition of an experiment hits its ideal significant solutions. More specifically, if $P$ is a distribution of a state vector of $n$ qubits and $\overline{p}$ is the median, then the \emph{heavy output} is $H:=\{x \mid P(x) > \overline{p}\}$. Now let $\{q(x) \mid x \in \{0,1\}^n\}$ be the probabilities of a noisy experiment, then the \emph{heavy output probability} is $\sum_{x: x \in H} q(x)$, i.e. the probability that the noisy experiment hits the heavy output.



We use the \texttt{AerSimulator} to compute \gls{HOP} for the benchmark circuits. For the ideal distribution we run the transpiled circuit using the \texttt{statevector} method. The noise model was generated with the same backend used to generate the circuits, i.e. the \texttt{ibm\_kyoto} hardware. For the noisy experiment we apply the \texttt{densitymatrix} method with $k=1024$ shots and compute the \gls{HOP} value according to the definition above.

\subsection{Comparison of layer free optimization and optimization with layers}

\begin{figure}[ht!]
    \begin{center}
    \begin{subfigure}[c]{0.3\textwidth}
        \centering
    \begin{tikzpicture}[scale=0.6]
        \pgfmathsetmacro{\xmax}{800} 
    \pgfmathsetmacro{\ymax}{800} 
    \pgfmathsetmacro{\xmin}{100}     
    \pgfmathsetmacro{\ymin}{100}     
    \pgfmathsetmacro{\xtickinterval}{(\xmax - \xmin) / 5} 
    \pgfmathsetmacro{\ytickinterval}{(\ymax - \ymin) / 5} 
    \begin{axis}[axis lines=middle, xlabel={Non Layered}, ylabel={Layered}, xlabel style={align=center}, title=Linear / depth / depth,
                 xmin=\xmin, xmax=\xmax, 
                 ymin=\ymin, ymax=\ymax, 
                 xtick={\xmin,\xmin+\xtickinterval,\xmin+2*\xtickinterval,\xmin+3*\xtickinterval,\xmin+4*\xtickinterval,\xmax}, 
                 ytick={\ymin,\ymin+\ytickinterval,\ymin+2*\ytickinterval,\ymin+3*\ytickinterval,\ymin+4*\ytickinterval,\ymax}, 
                 tick label style={font=\small}] 

        \addplot[only marks, mark=*, mark size=1pt] table [col sep=semicolon, x index=0, y index=1] 
        {data/depth_ALG-LINEAR-DEPTH-NOLAYER_vs_ALG-LINEAR-DEPTH-LAYER.csv};
        
         \addplot[red, dashed, thick, domain=\xmin:\xmax] {x};
        
        
    \end{axis}
\end{tikzpicture}
        \caption{}
        \label{fig:depthdepthlinear}
    \end{subfigure}
    ~
    \begin{subfigure}[c]{0.3\textwidth}
        \centering
    \begin{tikzpicture}[scale=0.6]
        \pgfmathsetmacro{\xmax}{650} 
    \pgfmathsetmacro{\ymax}{650} 
    \pgfmathsetmacro{\xmin}{100}     
    \pgfmathsetmacro{\ymin}{100}     
    \pgfmathsetmacro{\xtickinterval}{(\xmax - \xmin) / 5} 
    \pgfmathsetmacro{\ytickinterval}{(\ymax - \ymin) / 5} 
    \begin{axis}[axis lines=middle, xlabel={Non Layered}, ylabel={Layered}, xlabel style={align=center}, title=Grid / depth / depth,
                 xmin=\xmin, xmax=\xmax, 
                 ymin=\ymin, ymax=\ymax, 
                 xtick={\xmin,\xmin+\xtickinterval,\xmin+2*\xtickinterval,\xmin+3*\xtickinterval,\xmin+4*\xtickinterval,\xmax}, 
                 ytick={\ymin,\ymin+\ytickinterval,\ymin+2*\ytickinterval,\ymin+3*\ytickinterval,\ymin+4*\ytickinterval,\ymax}, 
                 tick label style={font=\small}] 

        \addplot[only marks, mark=*, mark size=1pt] table [col sep=semicolon, x index=0, y index=1] 
        {data/depth_ALG-GRID-DEPTH-NOLAYER_vs_ALG-GRID-DEPTH-LAYER.csv};
        
         \addplot[red, dashed, thick, domain=\xmin:\xmax] {x};
        
        
    \end{axis}
\end{tikzpicture}
        \caption{}
        \label{fig:depthdepthgrid}
    \end{subfigure}
    ~
        \begin{subfigure}[c]{0.3\textwidth}
        \centering
    \begin{tikzpicture}[scale=0.6]
        \pgfmathsetmacro{\xmax}{1100} 
    \pgfmathsetmacro{\ymax}{1100} 
    \pgfmathsetmacro{\xmin}{100}     
    \pgfmathsetmacro{\ymin}{100}     
    \pgfmathsetmacro{\xtickinterval}{(\xmax - \xmin) / 5} 
    \pgfmathsetmacro{\ytickinterval}{(\ymax - \ymin) / 5} 
    \begin{axis}[axis lines=middle, xlabel={Non Layered}, ylabel={Layered}, xlabel style={align=center}, title=Y / depth / depth,
                 xmin=\xmin, xmax=\xmax, 
                 ymin=\ymin, ymax=\ymax, 
                 xtick={\xmin,\xmin+\xtickinterval,\xmin+2*\xtickinterval,\xmin+3*\xtickinterval,\xmin+4*\xtickinterval,\xmax}, 
                 ytick={\ymin,\ymin+\ytickinterval,\ymin+2*\ytickinterval,\ymin+3*\ytickinterval,\ymin+4*\ytickinterval,\ymax}, 
                 tick label style={font=\small}] 

        \addplot[only marks, mark=*, mark size=1pt] table [col sep=semicolon, x index=0, y index=1] 
        {data/depth_ALG-Y-DEPTH-NOLAYER_vs_ALG-Y-DEPTH-LAYER.csv};
        
         \addplot[red, dashed, thick, domain=\xmin:\xmax] {x};
        
        
    \end{axis}
\end{tikzpicture}
        \caption{}
        \label{fig:depthdepthy}
    \end{subfigure}

       \begin{subfigure}[c]{0.3\textwidth}
        \centering
    \begin{tikzpicture}[scale=0.6]
    \pgfmathsetmacro{\xmax}{42} 
    \pgfmathsetmacro{\ymax}{42} 
    \pgfmathsetmacro{\xmin}{0}     
    \pgfmathsetmacro{\ymin}{0}     
    \pgfmathsetmacro{\xtickinterval}{(\xmax - \xmin) / 5} 
    \pgfmathsetmacro{\ytickinterval}{(\ymax - \ymin) / 5} 
    \begin{axis}[axis lines=middle, xlabel={Non Layered}, ylabel={Layered}, xlabel style={align=center}, title=Linear / depth / num swaps,
                 xmin=\xmin, xmax=\xmax, 
                 ymin=\ymin, ymax=\ymax, 
                 xtick={\xmin,\xmin+\xtickinterval,\xmin+2*\xtickinterval,\xmin+3*\xtickinterval,\xmin+4*\xtickinterval,\xmax}, 
                 ytick={\ymin,\ymin+\ytickinterval,\ymin+2*\ytickinterval,\ymin+3*\ytickinterval,\ymin+4*\ytickinterval,\ymax}, 
                 tick label style={font=\small}] 

        \addplot[only marks, mark=*, mark size=1pt] table [col sep=semicolon, x index=0, y index=1] 
        {data/number_swaps_ALG-LINEAR-DEPTH-NOLAYER_vs_ALG-LINEAR-DEPTH-LAYER.csv};
        
         \addplot[red, dashed, thick, domain=\xmin:\xmax] {x};
        
        
    \end{axis}
\end{tikzpicture}
        \caption{}
        \label{fig:swapsdepthlinear}
    \end{subfigure}
    ~
    \begin{subfigure}[c]{0.3\textwidth}
        \centering
    \begin{tikzpicture}[scale=0.6]
    \pgfmathsetmacro{\xmax}{23} 
    \pgfmathsetmacro{\ymax}{23} 
    \pgfmathsetmacro{\xmin}{0}     
    \pgfmathsetmacro{\ymin}{0}     
    \pgfmathsetmacro{\xtickinterval}{(\xmax - \xmin) / 5} 
    \pgfmathsetmacro{\ytickinterval}{(\ymax - \ymin) / 5} 
    \begin{axis}[axis lines=middle, xlabel={Non Layered}, ylabel={Layered}, xlabel style={align=center}, title=Grid / depth / num swaps,
                 xmin=\xmin, xmax=\xmax, 
                 ymin=\ymin, ymax=\ymax, 
                 xtick={\xmin,\xmin+\xtickinterval,\xmin+2*\xtickinterval,\xmin+3*\xtickinterval,\xmin+4*\xtickinterval,\xmax}, 
                 ytick={\ymin,\ymin+\ytickinterval,\ymin+2*\ytickinterval,\ymin+3*\ytickinterval,\ymin+4*\ytickinterval,\ymax}, 
                 tick label style={font=\small}] 

        \addplot[only marks, mark=*, mark size=1pt] table [col sep=semicolon, x index=0, y index=1] 
        {data/number_swaps_ALG-GRID-DEPTH-NOLAYER_vs_ALG-GRID-DEPTH-LAYER.csv};
        
         \addplot[red, dashed, thick, domain=\xmin:\xmax] {x};
        
        
    \end{axis}
\end{tikzpicture}
        \caption{}
        \label{fig:swapsdepthgrid}
    \end{subfigure}
    ~
        \begin{subfigure}[c]{0.3\textwidth}
        \centering
    \begin{tikzpicture}[scale=0.6]
    \pgfmathsetmacro{\xmax}{35} 
    \pgfmathsetmacro{\ymax}{35} 
    \pgfmathsetmacro{\xmin}{0}     
    \pgfmathsetmacro{\ymin}{0}     
    \pgfmathsetmacro{\xtickinterval}{(\xmax - \xmin) / 5} 
    \pgfmathsetmacro{\ytickinterval}{(\ymax - \ymin) / 5} 
    \begin{axis}[axis lines=middle, xlabel={Non Layered}, ylabel={Layered}, xlabel style={align=center}, title=Y / depth / num swaps,
                 xmin=\xmin, xmax=\xmax, 
                 ymin=\ymin, ymax=\ymax, 
                 xtick={\xmin,\xmin+\xtickinterval,\xmin+2*\xtickinterval,\xmin+3*\xtickinterval,\xmin+4*\xtickinterval,\xmax}, 
                 ytick={\ymin,\ymin+\ytickinterval,\ymin+2*\ytickinterval,\ymin+3*\ytickinterval,\ymin+4*\ytickinterval,\ymax}, 
                 tick label style={font=\small}] 

        \addplot[only marks, mark=*, mark size=1pt] table [col sep=semicolon, x index=0, y index=1] 
        {data/number_swaps_ALG-Y-DEPTH-NOLAYER_vs_ALG-Y-DEPTH-LAYER.csv};
        
         \addplot[red, dashed, thick, domain=\xmin:\xmax] {x};
        
        
    \end{axis}
\end{tikzpicture}
        \caption{}
        \label{fig:swapsdepthy}
    \end{subfigure}

        \begin{subfigure}[c]{0.3\textwidth}
        \centering
    \begin{tikzpicture}[scale=0.6]
        \pgfmathsetmacro{\xmax}{500} 
    \pgfmathsetmacro{\ymax}{500} 
    \pgfmathsetmacro{\xmin}{50}     
    \pgfmathsetmacro{\ymin}{50}     
    \pgfmathsetmacro{\xtickinterval}{(\xmax - \xmin) / 5} 
    \pgfmathsetmacro{\ytickinterval}{(\ymax - \ymin) / 5} 
    \begin{axis}[axis lines=middle, xlabel={Non Layered}, ylabel={Layered}, xlabel style={align=center}, title= Linear / depth / unweighted depth,
                 xmin=\xmin, xmax=\xmax, 
                 ymin=\ymin, ymax=\ymax, 
                 xtick={\xmin,\xmin+\xtickinterval,\xmin+2*\xtickinterval,\xmin+3*\xtickinterval,\xmin+4*\xtickinterval,\xmax}, 
                 ytick={\ymin,\ymin+\ytickinterval,\ymin+2*\ytickinterval,\ymin+3*\ytickinterval,\ymin+4*\ytickinterval,\ymax}, 
                 tick label style={font=\small}] 

        \addplot[only marks, mark=*, mark size=1pt] table [col sep=semicolon, x index=0, y index=1] 
        {data/unweighteddepth_ALG-LINEAR-DEPTH-NOLAYER_vs_ALG-LINEAR-DEPTH-LAYER.csv};
        
         \addplot[red, dashed, thick, domain=\xmin:\xmax] {x};
        
        
    \end{axis}
\end{tikzpicture}
        \caption{}
        \label{fig:unweighteddepthdepthlinear}
    \end{subfigure}
    ~
    \begin{subfigure}[c]{0.3\textwidth}
        \centering
    \begin{tikzpicture}[scale=0.6]
        \pgfmathsetmacro{\xmax}{450} 
    \pgfmathsetmacro{\ymax}{450} 
    \pgfmathsetmacro{\xmin}{50}     
    \pgfmathsetmacro{\ymin}{50}     
    \pgfmathsetmacro{\xtickinterval}{(\xmax - \xmin) / 5} 
    \pgfmathsetmacro{\ytickinterval}{(\ymax - \ymin) / 5} 
    \begin{axis}[axis lines=middle, xlabel={Non Layered}, ylabel={Layered}, xlabel style={align=center}, title=Grid / depth / unweighted depth,
                 xmin=\xmin, xmax=\xmax, 
                 ymin=\ymin, ymax=\ymax, 
                 xtick={\xmin,\xmin+\xtickinterval,\xmin+2*\xtickinterval,\xmin+3*\xtickinterval,\xmin+4*\xtickinterval,\xmax}, 
                 ytick={\ymin,\ymin+\ytickinterval,\ymin+2*\ytickinterval,\ymin+3*\ytickinterval,\ymin+4*\ytickinterval,\ymax}, 
                 tick label style={font=\small}] 

        \addplot[only marks, mark=*, mark size=1pt] table [col sep=semicolon, x index=0, y index=1] 
        {data/unweighteddepth_ALG-GRID-DEPTH-NOLAYER_vs_ALG-GRID-DEPTH-LAYER.csv};
        
         \addplot[red, dashed, thick, domain=\xmin:\xmax] {x};
        
        
    \end{axis}
\end{tikzpicture}
        \caption{}
        \label{fig:unweighteddepthdepthgrid}
    \end{subfigure}
    ~
        \begin{subfigure}[c]{0.3\textwidth}
        \centering
    \begin{tikzpicture}[scale=0.6]
        \pgfmathsetmacro{\xmax}{550} 
    \pgfmathsetmacro{\ymax}{550} 
    \pgfmathsetmacro{\xmin}{50}     
    \pgfmathsetmacro{\ymin}{50}     
    \pgfmathsetmacro{\xtickinterval}{(\xmax - \xmin) / 5} 
    \pgfmathsetmacro{\ytickinterval}{(\ymax - \ymin) / 5} 
    \begin{axis}[axis lines=middle, xlabel={Non Layered}, ylabel={Layered}, xlabel style={align=center}, title=Y / depth / unweighted depth,
                 xmin=\xmin, xmax=\xmax, 
                 ymin=\ymin, ymax=\ymax, 
                 xtick={\xmin,\xmin+\xtickinterval,\xmin+2*\xtickinterval,\xmin+3*\xtickinterval,\xmin+4*\xtickinterval,\xmax}, 
                 ytick={\ymin,\ymin+\ytickinterval,\ymin+2*\ytickinterval,\ymin+3*\ytickinterval,\ymin+4*\ytickinterval,\ymax}, 
                 tick label style={font=\small}] 

        \addplot[only marks, mark=*, mark size=1pt] table [col sep=semicolon, x index=0, y index=1] 
        {data/unweighteddepth_ALG-Y-DEPTH-NOLAYER_vs_ALG-Y-DEPTH-LAYER.csv};
        
         \addplot[red, dashed, thick, domain=\xmin:\xmax] {x};
        
        
    \end{axis}
\end{tikzpicture}
        \caption{}
        \label{fig:unweighteddepthdepthy}
    \end{subfigure}
    \caption{Parity plots for the depth objective for the three hardware graphs Linear, Grid and Y. Comparison is done with regard to the depth, the number of SWAPs and the unweighted depth.}
    \label{fig:paritydepth}
    \end{center}
\end{figure}
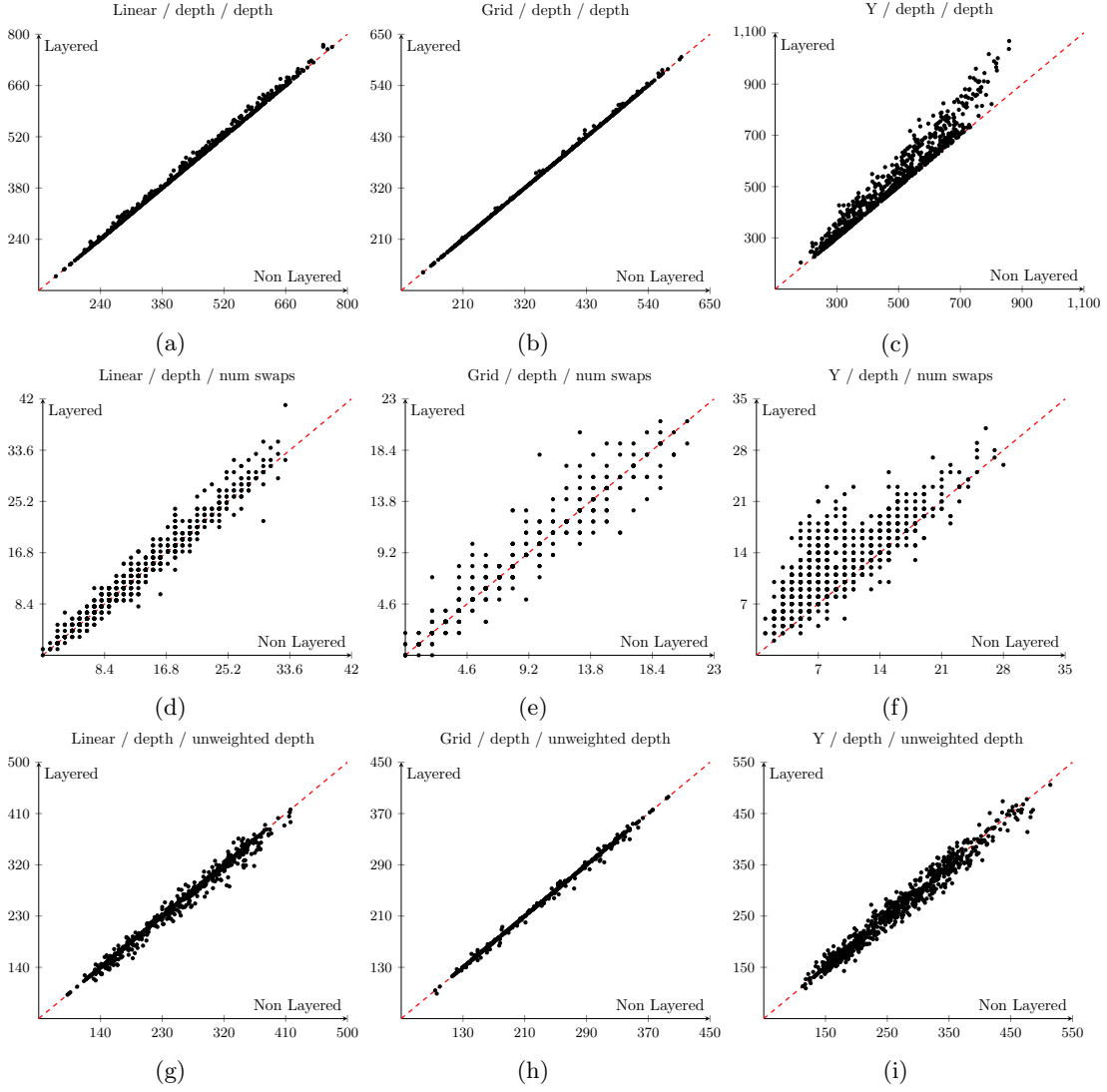

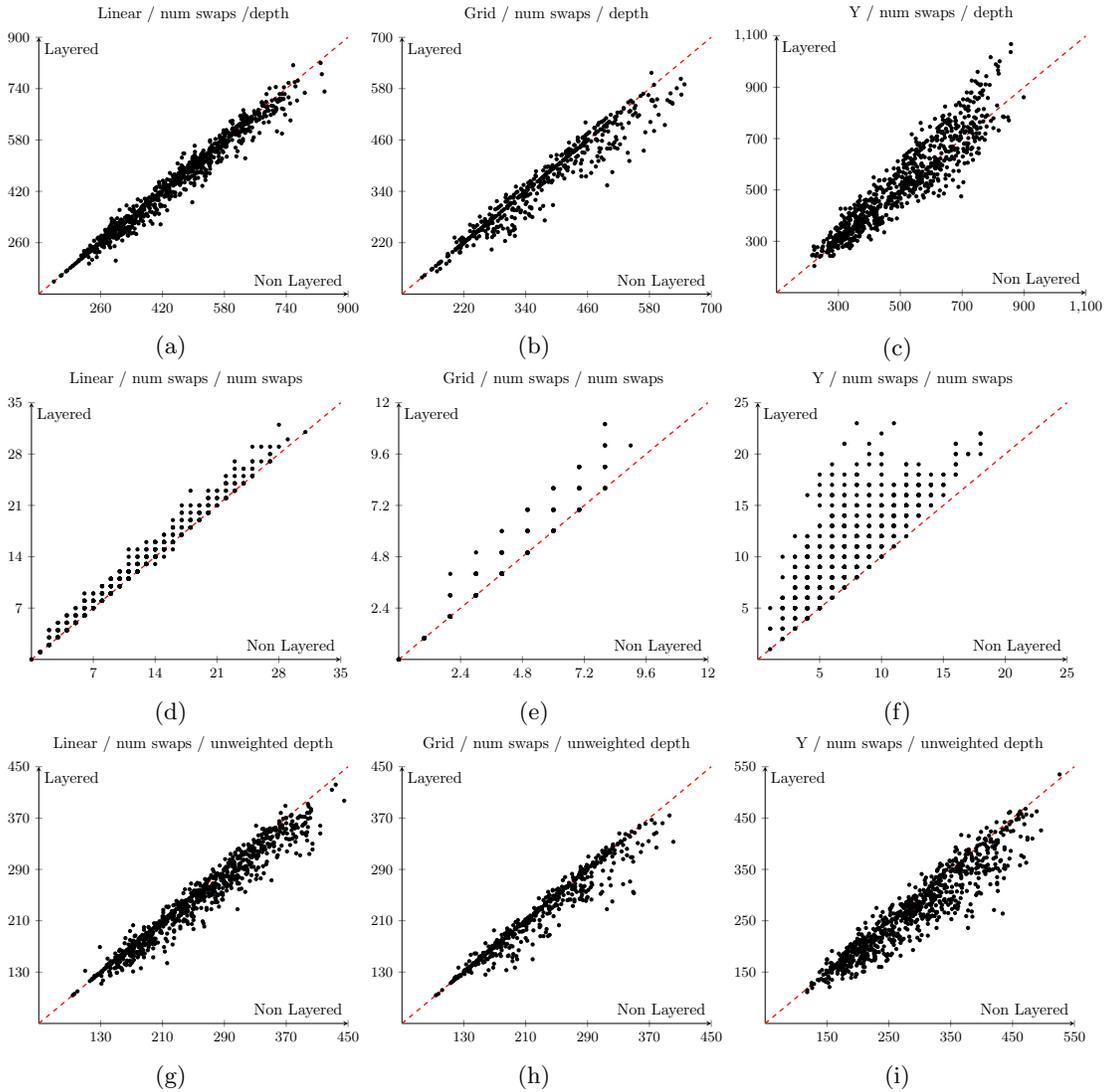
\begin{figure}[ht!]
    \begin{center}
    \begin{subfigure}[c]{0.3\textwidth}
        \centering
    \begin{tikzpicture}[scale=0.6]
        \pgfmathsetmacro{\xmax}{900} 
    \pgfmathsetmacro{\ymax}{900} 
    \pgfmathsetmacro{\xmin}{100}     
    \pgfmathsetmacro{\ymin}{100}     
    \pgfmathsetmacro{\xtickinterval}{(\xmax - \xmin) / 5} 
    \pgfmathsetmacro{\ytickinterval}{(\ymax - \ymin) / 5} 
    \begin{axis}[axis lines=middle, xlabel={Non Layered}, ylabel={Layered}, xlabel style={align=center}, title=Linear / num swaps /depth ,
                 xmin=\xmin, xmax=\xmax, 
                 ymin=\ymin, ymax=\ymax, 
                 xtick={\xmin,\xmin+\xtickinterval,\xmin+2*\xtickinterval,\xmin+3*\xtickinterval,\xmin+4*\xtickinterval,\xmax}, 
                 ytick={\ymin,\ymin+\ytickinterval,\ymin+2*\ytickinterval,\ymin+3*\ytickinterval,\ymin+4*\ytickinterval,\ymax}, 
                 tick label style={font=\small}] 

        \addplot[only marks, mark=*, mark size=1pt] table [col sep=semicolon, x index=0, y index=1] 
        {data/depth_ALG-LINEAR-SWAPS-NOLAYER_vs_ALG-LINEAR-SWAPS-LAYER.csv};
        
         \addplot[red, dashed, thick, domain=\xmin:\xmax] {x};
        
        
    \end{axis}
\end{tikzpicture}
        \caption{}
        \label{fig:depthswapslinear}
    \end{subfigure}
    ~
    \begin{subfigure}[c]{0.3\textwidth}
        \centering
    \begin{tikzpicture}[scale=0.6]
        \pgfmathsetmacro{\xmax}{700} 
    \pgfmathsetmacro{\ymax}{700} 
    \pgfmathsetmacro{\xmin}{100}     
    \pgfmathsetmacro{\ymin}{100}     
    \pgfmathsetmacro{\xtickinterval}{(\xmax - \xmin) / 5} 
    \pgfmathsetmacro{\ytickinterval}{(\ymax - \ymin) / 5} 
    \begin{axis}[axis lines=middle, xlabel={Non Layered}, ylabel={Layered}, xlabel style={align=center}, title=Grid / num swaps / depth ,
                 xmin=\xmin, xmax=\xmax, 
                 ymin=\ymin, ymax=\ymax, 
                 xtick={\xmin,\xmin+\xtickinterval,\xmin+2*\xtickinterval,\xmin+3*\xtickinterval,\xmin+4*\xtickinterval,\xmax}, 
                 ytick={\ymin,\ymin+\ytickinterval,\ymin+2*\ytickinterval,\ymin+3*\ytickinterval,\ymin+4*\ytickinterval,\ymax}, 
                 tick label style={font=\small}] 

        \addplot[only marks, mark=*, mark size=1pt] table [col sep=semicolon, x index=0, y index=1] 
        {data/depth_ALG-GRID-SWAPS-NOLAYER_vs_ALG-GRID-SWAPS-LAYER.csv};
        
         \addplot[red, dashed, thick, domain=\xmin:\xmax] {x};
        
        
    \end{axis}
\end{tikzpicture}
        \caption{}
        \label{fig:depthswapsgrid}
    \end{subfigure}
    ~
        \begin{subfigure}[c]{0.3\textwidth}
        \centering
    \begin{tikzpicture}[scale=0.6]
        \pgfmathsetmacro{\xmax}{1100} 
    \pgfmathsetmacro{\ymax}{1100} 
    \pgfmathsetmacro{\xmin}{100}     
    \pgfmathsetmacro{\ymin}{100}     
    \pgfmathsetmacro{\xtickinterval}{(\xmax - \xmin) / 5} 
    \pgfmathsetmacro{\ytickinterval}{(\ymax - \ymin) / 5} 
    \begin{axis}[axis lines=middle, xlabel={Non Layered}, ylabel={Layered}, xlabel style={align=center}, title=Y / num swaps / depth,
                 xmin=\xmin, xmax=\xmax, 
                 ymin=\ymin, ymax=\ymax, 
                 xtick={\xmin,\xmin+\xtickinterval,\xmin+2*\xtickinterval,\xmin+3*\xtickinterval,\xmin+4*\xtickinterval,\xmax}, 
                 ytick={\ymin,\ymin+\ytickinterval,\ymin+2*\ytickinterval,\ymin+3*\ytickinterval,\ymin+4*\ytickinterval,\ymax}, 
                 tick label style={font=\small}] 

        \addplot[only marks, mark=*, mark size=1pt] table [col sep=semicolon, x index=0, y index=1] 
        {data/depth_ALG-Y-SWAPS-NOLAYER_vs_ALG-Y-SWAPS-LAYER.csv};
        
         \addplot[red, dashed, thick, domain=\xmin:\xmax] {x};
        
        
    \end{axis}
\end{tikzpicture}
        \caption{}
        \label{fig:depthswapsy}
    \end{subfigure}

       \begin{subfigure}[c]{0.3\textwidth}
        \centering
    \begin{tikzpicture}[scale=0.6]
    \pgfmathsetmacro{\xmax}{35} 
    \pgfmathsetmacro{\ymax}{35} 
    \pgfmathsetmacro{\xmin}{0}     
    \pgfmathsetmacro{\ymin}{0}     
    \pgfmathsetmacro{\xtickinterval}{(\xmax - \xmin) / 5} 
    \pgfmathsetmacro{\ytickinterval}{(\ymax - \ymin) / 5} 
    \begin{axis}[axis lines=middle, xlabel={Non Layered}, ylabel={Layered}, xlabel style={align=center}, title=Linear / num swaps / num swaps,
                 xmin=\xmin, xmax=\xmax, 
                 ymin=\ymin, ymax=\ymax, 
                 xtick={\xmin,\xmin+\xtickinterval,\xmin+2*\xtickinterval,\xmin+3*\xtickinterval,\xmin+4*\xtickinterval,\xmax}, 
                 ytick={\ymin,\ymin+\ytickinterval,\ymin+2*\ytickinterval,\ymin+3*\ytickinterval,\ymin+4*\ytickinterval,\ymax}, 
                 tick label style={font=\small}] 

        \addplot[only marks, mark=*, mark size=1pt] table [col sep=semicolon, x index=0, y index=1] 
        {data/number_swaps_ALG-LINEAR-SWAPS-NOLAYER_vs_ALG-LINEAR-SWAPS-LAYER.csv};
        
         \addplot[red, dashed, thick, domain=\xmin:\xmax] {x};
        
        
    \end{axis}
\end{tikzpicture}
        \caption{}
        \label{fig:swapsswapslinear}
    \end{subfigure}
    ~
    \begin{subfigure}[c]{0.3\textwidth}
        \centering
    \begin{tikzpicture}[scale=0.6]
    \pgfmathsetmacro{\xmax}{12} 
    \pgfmathsetmacro{\ymax}{12} 
    \pgfmathsetmacro{\xmin}{0}     
    \pgfmathsetmacro{\ymin}{0}     
    \pgfmathsetmacro{\xtickinterval}{(\xmax - \xmin) / 5} 
    \pgfmathsetmacro{\ytickinterval}{(\ymax - \ymin) / 5} 
    \begin{axis}[axis lines=middle, xlabel={Non Layered}, ylabel={Layered}, xlabel style={align=center}, title=Grid / num swaps / num swaps,
                 xmin=\xmin, xmax=\xmax, 
                 ymin=\ymin, ymax=\ymax, 
                 xtick={\xmin,\xmin+\xtickinterval,\xmin+2*\xtickinterval,\xmin+3*\xtickinterval,\xmin+4*\xtickinterval,\xmax}, 
                 ytick={\ymin,\ymin+\ytickinterval,\ymin+2*\ytickinterval,\ymin+3*\ytickinterval,\ymin+4*\ytickinterval,\ymax}, 
                 tick label style={font=\small}] 

        \addplot[only marks, mark=*, mark size=1pt] table [col sep=semicolon, x index=0, y index=1] 
        {data/number_swaps_ALG-GRID-SWAPS-NOLAYER_vs_ALG-GRID-SWAPS-LAYER.csv};
        
         \addplot[red, dashed, thick, domain=\xmin:\xmax] {x};
        
        
    \end{axis}
\end{tikzpicture}
        \caption{}
        \label{fig:swapsswapsgrid}
    \end{subfigure}
    ~
        \begin{subfigure}[c]{0.3\textwidth}
        \centering
    \begin{tikzpicture}[scale=0.6]
    \pgfmathsetmacro{\xmax}{25} 
    \pgfmathsetmacro{\ymax}{25} 
    \pgfmathsetmacro{\xmin}{0}     
    \pgfmathsetmacro{\ymin}{0}     
    \pgfmathsetmacro{\xtickinterval}{(\xmax - \xmin) / 5} 
    \pgfmathsetmacro{\ytickinterval}{(\ymax - \ymin) / 5} 
    \begin{axis}[axis lines=middle, xlabel={Non Layered}, ylabel={Layered}, xlabel style={align=center}, title=Y / num swaps / num swaps,
                 xmin=\xmin, xmax=\xmax, 
                 ymin=\ymin, ymax=\ymax, 
                 xtick={\xmin,\xmin+\xtickinterval,\xmin+2*\xtickinterval,\xmin+3*\xtickinterval,\xmin+4*\xtickinterval,\xmax}, 
                 ytick={\ymin,\ymin+\ytickinterval,\ymin+2*\ytickinterval,\ymin+3*\ytickinterval,\ymin+4*\ytickinterval,\ymax}, 
                 tick label style={font=\small}] 

        \addplot[only marks, mark=*, mark size=1pt] table [col sep=semicolon, x index=0, y index=1] 
        {data/number_swaps_ALG-Y-SWAPS-NOLAYER_vs_ALG-Y-SWAPS-LAYER.csv};
        
         \addplot[red, dashed, thick, domain=\xmin:\xmax] {x};
        
        
    \end{axis}
\end{tikzpicture}
        \caption{}
        \label{fig:swapsswapsy}
    \end{subfigure}

        \begin{subfigure}[c]{0.3\textwidth}
        \centering
    \begin{tikzpicture}[scale=0.6]
        \pgfmathsetmacro{\xmax}{450} 
    \pgfmathsetmacro{\ymax}{450} 
    \pgfmathsetmacro{\xmin}{50}     
    \pgfmathsetmacro{\ymin}{50}     
    \pgfmathsetmacro{\xtickinterval}{(\xmax - \xmin) / 5} 
    \pgfmathsetmacro{\ytickinterval}{(\ymax - \ymin) / 5} 
    \begin{axis}[axis lines=middle, xlabel={Non Layered}, ylabel={Layered}, xlabel style={align=center}, title=Linear / num swaps / unweighted depth,
                 xmin=\xmin, xmax=\xmax, 
                 ymin=\ymin, ymax=\ymax, 
                 xtick={\xmin,\xmin+\xtickinterval,\xmin+2*\xtickinterval,\xmin+3*\xtickinterval,\xmin+4*\xtickinterval,\xmax}, 
                 ytick={\ymin,\ymin+\ytickinterval,\ymin+2*\ytickinterval,\ymin+3*\ytickinterval,\ymin+4*\ytickinterval,\ymax}, 
                 tick label style={font=\small}] 

        \addplot[only marks, mark=*, mark size=1pt] table [col sep=semicolon, x index=0, y index=1] 
        {data/unweighteddepth_ALG-LINEAR-SWAPS-NOLAYER_vs_ALG-LINEAR-SWAPS-LAYER.csv};
        
         \addplot[red, dashed, thick, domain=\xmin:\xmax] {x};
        
        
    \end{axis}
\end{tikzpicture}
        \caption{}
        \label{fig:unweighteddepthswapslinear}
    \end{subfigure}
    ~
    \begin{subfigure}[c]{0.3\textwidth}
        \centering
    \begin{tikzpicture}[scale=0.6]
        \pgfmathsetmacro{\xmax}{450} 
    \pgfmathsetmacro{\ymax}{450} 
    \pgfmathsetmacro{\xmin}{50}     
    \pgfmathsetmacro{\ymin}{50}     
    \pgfmathsetmacro{\xtickinterval}{(\xmax - \xmin) / 5} 
    \pgfmathsetmacro{\ytickinterval}{(\ymax - \ymin) / 5} 
    \begin{axis}[axis lines=middle, xlabel={Non Layered}, ylabel={Layered}, xlabel style={align=center}, title=Grid / num swaps / unweighted depth,
                 xmin=\xmin, xmax=\xmax, 
                 ymin=\ymin, ymax=\ymax, 
                 xtick={\xmin,\xmin+\xtickinterval,\xmin+2*\xtickinterval,\xmin+3*\xtickinterval,\xmin+4*\xtickinterval,\xmax}, 
                 ytick={\ymin,\ymin+\ytickinterval,\ymin+2*\ytickinterval,\ymin+3*\ytickinterval,\ymin+4*\ytickinterval,\ymax}, 
                 tick label style={font=\small}] 

        \addplot[only marks, mark=*, mark size=1pt] table [col sep=semicolon, x index=0, y index=1] 
        {data/unweighteddepth_ALG-GRID-SWAPS-NOLAYER_vs_ALG-GRID-SWAPS-LAYER.csv};
        
         \addplot[red, dashed, thick, domain=\xmin:\xmax] {x};
        
        
    \end{axis}
\end{tikzpicture}
        \caption{}
        \label{fig:unweighteddepthswapsgrid}
    \end{subfigure}
    ~
        \begin{subfigure}[c]{0.3\textwidth}
        \centering
    \begin{tikzpicture}[scale=0.6]
        \pgfmathsetmacro{\xmax}{550} 
    \pgfmathsetmacro{\ymax}{550} 
    \pgfmathsetmacro{\xmin}{50}     
    \pgfmathsetmacro{\ymin}{50}     
    \pgfmathsetmacro{\xtickinterval}{(\xmax - \xmin) / 5} 
    \pgfmathsetmacro{\ytickinterval}{(\ymax - \ymin) / 5} 
    \begin{axis}[axis lines=middle, xlabel={Non Layered}, ylabel={Layered}, xlabel style={align=center}, title=Y / num swaps / unweighted depth,
                 xmin=\xmin, xmax=\xmax, 
                 ymin=\ymin, ymax=\ymax, 
                 xtick={\xmin,\xmin+\xtickinterval,\xmin+2*\xtickinterval,\xmin+3*\xtickinterval,\xmin+4*\xtickinterval,\xmax}, 
                 ytick={\ymin,\ymin+\ytickinterval,\ymin+2*\ytickinterval,\ymin+3*\ytickinterval,\ymin+4*\ytickinterval,\ymax}, 
                 tick label style={font=\small}] 

        \addplot[only marks, mark=*, mark size=1pt] table [col sep=semicolon, x index=0, y index=1] 
        {data/unweighteddepth_ALG-Y-SWAPS-NOLAYER_vs_ALG-Y-SWAPS-LAYER.csv};
        
         \addplot[red, dashed, thick, domain=\xmin:\xmax] {x};
        
        
    \end{axis}
\end{tikzpicture}
        \caption{}
        \label{fig:unweighteddepthswapsy}
    \end{subfigure}
    \caption{Parity plots for the number of SWAPs objective for the three hardware graphs Linear, Grid and Y. Comparison is done with regard to the depth, the number of SWAPs and the unweighted depth.}
    \label{fig:parityswaps}
    \end{center}
\end{figure}

We display the results for the two objectives and three metrics in parity plots. In these plots, we scatter the solution values of the computed instances, where the $x$-value corresponds to the metric value of the non-layered instance, and the $y$-value represents the value of the layered variant. The dashed red line marks the line of parity, indicating where both instance solutions agree on the considered metric. We only consider the instances where both the layered and non-layered approach converged after a given time limit of \SI{500}{\second}; all others are omitted. The comparison of results for the depth objective is presented in Figure \ref{fig:paritydepth}, while the results for the number of SWAPs are shown in Figure \ref{fig:parityswaps}. In both figures, the columns correspond to the hardware graphs (Linear, Grid, and Y), and the rows correspond to the metrics (depth, number of SWAPs, and unweighted depth). Additionally, each plot title identifies the combination of hardware graph, objective, and metric used. 

In Subfigures \ref{fig:depthdepthlinear} - \ref{fig:depthdepthy} and \ref{fig:swapsswapslinear} - \ref{fig:swapsswapsy} it can be seen that for all hardware graphs, the non-layered approach outperforms the layered one when comparing the metric used for optimization. The effect is most present for the Y graph instances, with over three more SWAPs on average and a depth longer than $37$ units of time. On the other hand, the non-layered approach provides the least advantage for the grid-graph instances with $0.36$ more SWAPs on average and an increase of $0.24$ in circuit duration. This finding is expected, given that the Grid graph has the strongest connectivity. Additionally the Y-graph has a central node; for a Y graph with $4$ nodes for example, a SWAP is expected in each layer with more than one gate in the layered approach. In between lie the instances with a Linear hardware graph, with a noticeable increase in SWAPs of $0.46$ and an average duration $2.22$ units longer than with the non-layered approach. Additionally \ref{fig:swapsdepthlinear}-\ref{fig:swapsdepthy} show that optimization for the depth also has a positive effect on the number of SWAPs, where the non-layered approach is favorable as well. This is mainly due to the long duration of the SWAP gate, which significantly increases the execution time of the circuit when present. On the other hand, optimization for the number of SWAPs does not seem to have a significant effect on the circuit depth as seen in \ref{fig:depthswapslinear} - \ref{fig:depthswapsy}, especially on Linear and Grid hardware graphs. Lastly there is little to no effect on the unweighted depth for both optimization targets as seen in \ref{fig:unweighteddepthdepthlinear} - \ref{fig:unweighteddepthdepthy} and \ref{fig:unweighteddepthswapslinear} - \ref{fig:unweighteddepthswapsy}. This highlights the importance of taking gate durations into account, when optimizing for the actual makespan of the circuit.

Additionally we take a closer look at the results of the metric comparison with the same optimization target in \ref{fig:depthdepthlinear} - \ref{fig:depthdepthy} and \ref{fig:swapsswapslinear} - \ref{fig:swapsswapsy}. As many solution values agree for the same instance in the parity plots, we additionally compute another metric, the \emph{relative mean deviation} RMD; Table \ref{tab:meanbias} shows these results.
The number of instances of each graph and objective combination, which are successfully solved before timeout is denoted by $N^S$. The number of instances where both the layered and non-layered approach returned the same objective value is denoted by $N^=$. The relative mean deviation is computed as the mean of the deviations of all solution values from the layered and non-layered instances: 

\[
\RMD = \frac{1}{N^S}\sum_{i} \frac{(y_i^L - y_i^{NL})}{y_i^L}
\]

where \(y_i^L\) and \(y_i^{NL}\) are the objective values from the layered and non-layered instances, respectively.

Since many instances for the Linear and Grid hardware graphs agree with respect to the metric used, we also compute the relative mean deviation for instances that did not yield the same value, denoted as $\RMD^{\neq}$. This metric is of interest as for the fact, that the instances have been solved to optimality by Algorithm \ref{alg:treesearch}. Therefore, instances where both approached agree on their objective value have no potential of optimization at all, when considering a non-layering vs. a layering approach.

\begin{table}[ht!]
    \centering
    \begin{tabular}{|c|c|c|c|c|c|c|}
    \hline 
         Hardware Graph    & Objective & $N$ & $N^S$ & $N^=$ & $\RMD$ & $\RMD^{\neq}$\\ 
        \hline
        \multirow{2}{*}{Linear} &  depth & \multirow{2}{*}{$900$}   &    $900$     &  $679$   & \ $0.48\%$  & \ $1.96\%$ \\ 
        \cline{2-2} \cline{4-7}
             & num SWAPs &   & $892$ & $604$ & \  $4.05\%$ & $12.53\%$ \\ 
            
        \hline
        \multirow{2}{*}{Grid} &  depth &    \multirow{2}{*}{$600$}         &    $595$  &  $554$   & \ $0.09\%$    & \ $1.26\%$\\ 
        \cline{2-2} \cline{4-7}
            
            & num SWAPs &   & $530$ &  $430$ & \ $3.92\%$ & $20.75\%$\\ 
        \hline
        \multirow{2}{*}{Y} & depth & \multirow{2}{*}{$900$}  &   $815$        & $152$  & \ $6.38\%$  & \ $7.85\%$\\ 
        \cline{2-2} \cline{4-7}
            
            & num SWAPs &   & $811$ & $59$& $30.79\%$ & $33.21\%$\\ 
        \hline
    \end{tabular} 
    \caption{Relative mean deviation comparison for \ref{fig:depthdepthlinear} - \ref{fig:depthdepthy} and \ref{fig:swapsswapslinear} - \ref{fig:swapsswapsy}.}
    \label{tab:meanbias}
\end{table}

Lastly we display the mean of the \gls{HOP} values of the instances computed and their difference in Table \ref{tab:hopvaluemean}. The \gls{HOP} value means range between $64\%$ and $71\%$, with differences in the sub-percentage area. This is because we do not explicitly optimize for the \gls{HOP} value; however, a positive effect is observed for the non-layered approach. The most noticeable difference of layered vs. non-layered approach can be seen in the Y-graph instances, as for the objective values in \ref{fig:depthdepthy} and \ref{fig:swapsswapsy}. The difference for the other hardware graphs on the other hand is small. This can be explained by the high agreement of solutions for Grid and Linear graphs, in contrast to the instances on Y-graphs. Altogether, except for the number of SWAPs objective for the Grid hardware graph, the non-layered approach dominates the layered one with regard to the \gls{HOP} mean.

\begin{table}[ht]
    \centering
    \begin{tabular}{|c|c|c|c|c|}
    \hline 
         Hardware Graph    & Objective & Non-Layered & Layered  & Difference \\ 
        \hline
        \multirow{2}{*}{Linear} &  Depth &   $ 64.13\%$  & $64.02\%$   & \ \   $0.11\%$\\ 
        \cline{2-5}
            & Swaps & $64.61\%$ & $64.42\%$ & \ \  $0.19\%$\\ 
        \hline
        \multirow{2}{*}{Grid} &  Depth &  $69.06\%$   &   $68.92\%$          &  \ \    $0.14\%$    \\
        \cline{2-5}
            & Swaps & $69.78\%$ & $69.84\%$ &  $-0.06\%$\\ 
        \hline
        \multirow{2}{*}{Y} & Depth & $70.04\%$     &  $69.41\%$  &   \ \  $0.63\%$  \\ 
        \cline{2-5}
            & Swaps  & $70.02\%$ & $69.81\%$ & \ \  $0.21\%$\\
        \hline
    \end{tabular}
    \caption{\gls{HOP} value mean}
    \label{tab:hopvaluemean}
\end{table}

\section{Conclusions and future work}
\label{sec:conclusion}
In this paper, we analyzed the effect of considering or ignoring layers when optimizing the qubit mapping problem with respect to the minimization of circuit depths or number of SWAPs while also considering the gate execution time. We developed a custom and flexible branch and bound algorithm to solve this problem, accompanied by a novel bounding function. We then showed that when the hardware graph is sparsely connected (such as the linear graph or the Y graph), the difference between considering and ignoring layers can be significant. 

The flexibility of the algorithm offers the opportunity to easily test and experiments with new ideas. Here are some examples:
\begin{itemize}
    \item Bounding heuristics: the bounding function proposed in this paper is so-called admissible, meaning that the algorithm is guaranteed to find the optimal solution (given enough time). Non admissible bounding functions would allow the algorithm to possibly cut more branches, speeding up the solution process but loosing the optimality guarantees.
    \item Selection heuristics: while choosing the node with the lowest bound has been shown to produce smaller trees when looking to prove optimality \cite{dechter1985generalized}, different selection functions can produce a more heuristic behavior. For example, by limiting the number of nodes at a certain depth of the search tree, one could implement a beam-search heuristic.
    \item Noisy qubits: the algorithm currently ignores the fact that some physical qubits can be more noisy than others and should be avoided as much as possible. Interestingly, this noise can also change over time, and that could also be taken into account.
    \item Objective functions: the number of SWAPs and the depth of the circuit are two common proxy functions to evaluate the quality of a compiled circuit. However, other objectives functions such as the minimization of the error rate (suggested in \cite{nannicini2022optimal}) could be interesting to explore.
\end{itemize}

\section*{Acknowledgment}
We would like to thank the Research Council of Norway for providing partial funding for this work through project number 332023.

\bibliographystyle{plain}
\bibliography{main}

\begin{thebibliography}{10}

\bibitem{10.5555/3135595.3135617}
Scott Aaronson and Lijie Chen.
\newblock Complexity-theoretic foundations of quantum supremacy experiments.
\newblock In {\em Proceedings of the 32nd Computational Complexity Conference}, CCC '17, Dagstuhl, DEU, 2017. Schloss Dagstuhl--Leibniz-Zentrum fuer Informatik.

\bibitem{bhattacharjee2017depth}
Debjyoti Bhattacharjee and Anupam Chattopadhyay.
\newblock Depth-optimal quantum circuit placement for arbitrary topologies.
\newblock {\em arXiv preprint arXiv:1703.08540}, 2017.

\bibitem{botea2018complexity}
Adi Botea, Akihiro Kishimoto, and Radu Marinescu.
\newblock On the complexity of quantum circuit compilation.
\newblock In {\em Proceedings of the International Symposium on Combinatorial Search}, volume~9, pages 138--142, 2018.

\bibitem{cardama2024quantum}
F~Javier Cardama, Jorge V{\'a}zquez-P{\'e}rez, Tom{\'a}s~F Pena, Juan~C Pichel, and Andr{\'e}s G{\'o}mez.
\newblock Quantum compilation process: A survey.
\newblock In {\em European Conference on Parallel Processing}, pages 100--112. Springer, 2024.

\bibitem{Chen2024benchmarkingtrapped}
Jwo-Sy Chen, Erik Nielsen, Matthew Ebert, Volkan Inlek, Kenneth Wright, Vandiver Chaplin, Andrii Maksymov, Eduardo P{\'{a}}ez, Amrit Poudel, Peter Maunz, and John Gamble.
\newblock Benchmarking a trapped-ion quantum computer with 30 qubits.
\newblock {\em {Quantum}}, 8:1516, November 2024.

\bibitem{correr2024characterizing}
Guilherme~Il{\'a}rio Correr, Ivan Medina, Pedro~C Azado, Alexandre Drinko, and Diogo~O Soares-Pinto.
\newblock Characterizing randomness in parameterized quantum circuits through expressibility and average entanglement.
\newblock {\em Quantum Science and Technology}, 10(1):015008, 2024.

\bibitem{PhysRevA.100.032328}
Andrew~W. Cross, Lev~S. Bishop, Sarah Sheldon, Paul~D. Nation, and Jay~M. Gambetta.
\newblock Validating quantum computers using randomized model circuits.
\newblock {\em Phys. Rev. A}, 100:032328, Sep 2019.

\bibitem{dechter1985generalized}
Rina Dechter and Judea Pearl.
\newblock Generalized best-first search strategies and the optimality of {A*}.
\newblock {\em Journal of the ACM (JACM)}, 32(3):505--536, 1985.

\bibitem{ge2024quantum}
Yan Ge, Wu~Wenjie, Chen Yuheng, Pan Kaisen, Lu~Xudong, Zhou Zixiang, Wang Yuhan, Wang Ruocheng, and Yan Junchi.
\newblock Quantum circuit synthesis and compilation optimization: Overview and prospects.
\newblock {\em arXiv preprint arXiv:2407.00736}, 2024.

\bibitem{hetenyi2024creating}
Bence Het{\'e}nyi and James~R Wootton.
\newblock Creating entangled logical qubits in the heavy-hex lattice with topological codes.
\newblock {\em PRX Quantum}, 5(4):040334, 2024.

\bibitem{li2019tackling}
Gushu Li, Yufei Ding, and Yuan Xie.
\newblock Tackling the qubit mapping problem for nisq-era quantum devices.
\newblock In {\em Proceedings of the twenty-fourth international conference on architectural support for programming languages and operating systems}, pages 1001--1014, 2019.

\bibitem{maronese2022quantum}
Marco Maronese, Lorenzo Moro, Lorenzo Rocutto, and Enrico Prati.
\newblock Quantum compiling.
\newblock In {\em Quantum Computing Environments}, pages 39--74. Springer, 2022.

\bibitem{mulderij2023polynomial}
Jesse Mulderij, Karen~I Aardal, Irina Chiscop, and Frank Phillipson.
\newblock A polynomial size model with implicit swap gate counting for exact qubit reordering.
\newblock In {\em International Conference on Computational Science}, pages 72--89. Springer, 2023.

\bibitem{nannicini2022optimal}
Giacomo Nannicini, Lev~S Bishop, Oktay G{\"u}nl{\"u}k, and Petar Jurcevic.
\newblock Optimal qubit assignment and routing via integer programming.
\newblock {\em ACM Transactions on Quantum Computing}, 4(1):1--31, 2022.

\bibitem{NielsenChuang2010}
Michael~A. Nielsen and Isaac~L. Chuang.
\newblock {\em Quantum Computation and Quantum Information}.
\newblock Cambridge University Press, 2010.

\bibitem{siraichi2018qubit}
Marcos~Yukio Siraichi, Vin{\'\i}cius Fernandes~dos Santos, Caroline Collange, and Fernando Magno~Quint{\~a}o Pereira.
\newblock Qubit allocation.
\newblock In {\em Proceedings of the 2018 international symposium on code generation and optimization}, pages 113--125, 2018.

\bibitem{wagner2022improving}
Friedrich Wagner, Andreas B{\"a}rmann, Frauke Liers, and Markus Weissenb{\"a}ck.
\newblock Improving quantum computation by optimized qubit routing.
\newblock {\em Journal of Optimization Theory and Applications}, 197(3):1161--1194, 2023.

\bibitem{wille2019mapping}
Robert Wille, Lukas Burgholzer, and Alwin Zulehner.
\newblock Mapping quantum circuits to ibm qx architectures using the minimal number of swap and h operations.
\newblock In {\em Proceedings of the 56th Annual Design Automation Conference 2019}, pages 1--6, 2019.

\bibitem{zhang2021time}
Chi Zhang, Ari~B Hayes, Longfei Qiu, Yuwei Jin, Yanhao Chen, and Eddy~Z Zhang.
\newblock Time-optimal qubit mapping.
\newblock In {\em Proceedings of the 26th ACM International Conference on Architectural Support for Programming Languages and Operating Systems}, pages 360--374, 2021.

\bibitem{zhu2020exact}
Pengcheng Zhu, Xueyun Cheng, and Zhijin Guan.
\newblock An exact qubit allocation approach for nisq architectures.
\newblock {\em Quantum Information Processing}, 19(11):391, 2020.

\bibitem{zulehner2019compiling}
Alwin Zulehner and Robert Wille.
\newblock Compiling su (4) quantum circuits to ibm qx architectures.
\newblock In {\em Proceedings of the 24th Asia and South Pacific design automation conference}, pages 185--190, 2019.

\end{thebibliography}

\end{document}